\documentclass[largeformat]{interact}

\usepackage{epstopdf}
\usepackage{pifont}
\usepackage{setspace}
\usepackage{pbox}

\usepackage[natbibapa,nodoi]{apacite}
\renewcommand\bibliographytypesize{\fontsize{10}{12}\selectfont}

\usepackage{subcaption}
\usepackage{rotating}
\usepackage[inline]{enumitem}
\usepackage{graphicx}
\usepackage{caption}
\usepackage{float}
\usepackage[T1]{fontenc}
\usepackage{hyperref}

\theoremstyle{plain}

\theoremstyle{definition}

\theoremstyle{remark}

\usepackage{longtable}
\usepackage{booktabs}
\usepackage{pdflscape}
\usepackage{multirow}
\usepackage{url}
\usepackage{threeparttable}

\usepackage{kbordermatrix}

\usepackage[labelsep=space]{caption}

\begin{document}
\fontsize{12}{14}\selectfont   

\articletype{ARTICLE}

\title{Specifying Composites in Structural Equation Modeling: Traditional, Recent and New Approaches}

\author{
\name{J{\"o}rg Henseler\textsuperscript{a,b} and Xi Yu\textsuperscript{a}\thanks{CONTACT X. Yu. Email: x.yu-1@utwente.nl} and Tamara Schamberger\textsuperscript{c} and Gregory R. Hancock\textsuperscript{d} and Florian Schuberth\textsuperscript{a}}
\affil{\textsuperscript{a}University of Twente, Drienerlolaan 5, 7522 NB Enschede, The Netherlands; \textsuperscript{b}Nova Information Management School, Universidade de Lisboa, Campus de Campolide, 1070-312 Lisboa, Portugal; \textsuperscript{c}Faculty of Business Administration and Economics, Bielefeld University, Universitätsstraße 25, Bielefeld, Germany; \textsuperscript{d}College of Education, University of Maryland,
College Park, MD 20742, USA}
}

\maketitle
ORCID of authors:
J{\"o}rg Henseler (0000-0002-9736-3048), Xi Yu (0000-0001-8981-5799), Tamara Schamberger (0000-0002-7845-784X), Gregory R. Hancock (0000-0002-6313-006X), Florian Schuberth (0000-0002-2110-9086).

Funding details: This work was supported by national funds through FCT (Fundação para a Ciência e a Tecnologia), under the project - UIDB/04152 - Centro de Investigação em Gestão de Informação (MagIC)/NOVA IMS.

Disclosure statement: The authors report there are no competing interests to declare. 

Data availability statement: The dataset are from \citet{burgoyne_hambrick_altmann_2019}, which can be accessed via the Open Science Framework: \url{https:// osf.io/ndquz/?view_only=87cd1ad87c9e4fabba3e21c609ce9a14}.

\newpage

\begin{abstract}
Composites, or linear combinations of variables, play an important role in multivariate behavioral research. They appear in the form of indices, inventories, formative constructs, parcels, and emergent variables. Although structural equation modeling is widely used to study relations between variables, current approaches to incorporating composites have one or more limitations. These limitations include not modeling composite creation, not employing weights as model parameters, not being able to estimate weights, not allowing for composites in endogenous model positions, and not being able to assess whether a composite fully transmits the effects of or on its components. To address these limitations, we propose two novel composite specifications. The first specification combines the refined H–O specification of composites with phantom variables and uses the inverse of the weights as free parameters. The second specification blends the H-O specification with the pseudo-indicator approach of Rose et al. and uses the weights of all but one indicator as free parameters. We demonstrate the applicability of these specifications using an empirical example. The paper concludes with recommendations for choosing among the available composite specifications, as well as suggestions for future research.
\end{abstract}

\begin{keywords}
Structural equation modeling; emergent variables; formative constructs; Henseler–Ogasawara specification; index; inventory; composite construct
\end{keywords}


\section*{Introduction} \label{sec:intro}

Structural equation modeling (SEM) is a flexible modeling framework comprising various techniques such as regression analysis, path analysis, and analysis of variance as special cases \citep{Graham2008}.
It is widely applied in multivariate behavioral research.
For example, SEM has been used to study the relations between COVID-19 threat appraisal, fear of COVID-19, trust in information sources, conspiracy beliefs, and their effects on preventive behaviors \citep{Surina2021}.
It has also been used to investigate how parental phubbing and parenting practices influence children’s social withdrawal and aggression \citep{wang_qiao_li_lei_2021}. 
Similarly, SEM has been applied to study the determinants of diabetes self-management behavior \citep{pan_zhang_wang_zhao_zhao_ding_li_miao_fan_2023}.

In SEM, composites, i.e., linear combinations of variables \citep[e.g.,][]{henseler_2021} are commonly encountered. 
For example, composites appear in the form of additive indices and inventories \citep{Schamberger2023a}.
Composites are also used to reduce model complexity because they allow researchers to summarize the effects of multiple variables \citep{grace_bollen_2008,Whitt1986,Heise1972}
Moreover, in the measurement of latent variables, composites appear in the form of item parcels, which can improve reliability and avoid model convergence issues \citep{Little2013a,Bandalos2002}.
Furthermore, the concept proxy framework has been proposed, suggesting that both common factors and composites can serve as proxies for the concept under investigation \citep{Rigdon2012}.\footnote{For a critical discussion of the concept proxy framework, the interested reader is referred to \citet{Cadogan2023,Rigdon2023d,Cadogan2023a}.}
Moreover, formative constructs, i.e., constructs that are assumed to be defined and composed of other variables have been suggested to be modeled as composites. 
Similarly, composite constructs \citep{Benitez2018b}, composite attributes \citep{Rasoolimanesh2023}, forged concepts \citep{Yu2021}, 
and aggregate constructs \citep{Edwards2001} can be modeled as composites.

In the existing literature, there are at least seven ways to incorporate composites in SEM\footnote{In our study, we follow the SEM framework proposed by \citet{Joereskog1970a}, see also \citet{bollen_1989}.}: 
(i) the two-step approach; (ii) the one-step approach \citep[e.g.,][]{grace_bollen_2008}; (iii) the multiple indicators multiple causes (MIMIC) approach \citep{Bagozzi_Fornell_Larcker_1981}; (iv) the composite factor model specification \citep{Henseler2014};
(v) the pseudo-indicator approach \citep{Rose_Wagner_Mayer_Nagengast_2019};
(vi) the Henseler–Ogasawara (H–O) specification \citep{Schuberth2023HO}; and (vii) the refined H–O specification \citep{yu_schuberth_henseler_2023}. 
Despite the variety of approaches to studying composites using SEM, each approach has its limitations.
For example, the two-step approach does not account for composite creation in the model, and the one-step approach only allows for the modeling of composites as predictor variables.
Of the seven approaches, the refined H–O specification is arguably the most flexible.
It can mimic the results of the other approaches to a large extent  and allows for flexible modeling of unknown-weight composites, i.e., composites whose weights are freely estimated. 
However, it lacks the ability to express weights within the model specification.
Consequently, weights are not intuitively or conveniently visible in this approach, including in the corresponding software outputs. 
Researchers must perform calculations to obtain estimated weights, which requires effort.
Additionally, this approach has limitations when studying fixed-weight composites because it is unclear how to constrain the parameters to fix the weights to specific values. 

To address the limitations of the refined H–O specification, we propose two novel specifications for incorporating composites in SEM: the phantom-variable H–O specification and the blended H–O specification. 
The former departs from the refined H–O specification with unit weights \citep{Schuberth2025}, but it is modified by replacing each composite component with a phantom variable, i.e., a latent variable with a single indicator.
As a result, the weights equal the inverse of the loadings of the phantom variables on their respective components. 
The second new model specification is the blended H–O specification. 
It combines the central ideas of the H–O specification (extracting as many composites as components from a set of components) and the pseudo-indicator approach (expressing the pseudo indicator as the difference between the composite and its remaining components). 
The blended H–O specification offers the highest degree of modeling flexibility because it uses both composite loadings and indicator weights as model parameters.

The remainder of this article is structured as follows: 
The next section 
provides a detailed overview of the traditional and more recent approaches to studying composites in SEM.
Following this discussion, 
we introduce the two new approaches: the phantom-variable and the blended H–O specifications. 
Additionally, 
we illustrate the new approaches with an empirical example. 
Finally, we conclude the paper by discussing
the new approaches.  

\section*{Extant approaches to studying composites in SEM} \label{sec:traditional}

This section provides a detailed overview of the approaches used to study composites in SEM. 
We focus only on approaches that are compatible with the LISREL framework to SEM  \citet{Joereskog1970a}.
We begin by presenting the three traditional approaches to incorporating composites into SEM. 
First, we
elaborate on the two-step approach, which separates composite creation and model estimation into separate steps.
Second, 
we presents the one-step approach, which models the composite at the structural level. 
Third, we 
illustrate the MIMIC approach, which was proposed to emulate canonical correlation analysis within SEM.
Additionally, 
we discuss the composite factor model specification, which is equivalent to a common factor model with freely correlated unique terms.
Furthermore, we discuss three more recent approaches. 
Particularly, we discuss the pseudo-indicator approach, which models a composite using one of its components as a pseudo indicator.
Finally, 
we discuss the H–O specification and its refinement.
These specifications extract as many composites as components from a set of components and express their relations in terms of loadings rather than weights. 

To better illustrate the different approaches, we consider a hypothetical scenario in which a researcher aims to study the relations between two predictor composites $\eta_{1}$ and $\eta_{2}$ and an outcome composite $\eta_{3}$. 
The composite $\eta_{1}$ is composed of the observed variables $x_{11}$ to $x_{13}$, the composite $\eta_{2}$ is composed of the observed variables $x_{21}$ to $x_{23}$, and the composite $\eta_{3}$ is composed of the observed variables $x_{31}$ to $x_{34}$.

\subsection*{Two-step approach} 
The \textit{two-step approach} is commonly used to study composites \cite[see e.g.,][]{salehi2020,little2024,ahorsu2020}. 
In its first step, the two-step approach creates composite scores outside of SEM. 
Typically, a composite score is formed by summing the components. 
However, other types of fixed-weight composites are also possible; for example, the average of the components can be used. 
In the second step, the composite scores created in the first step are used as input for SEM to estimate the relations between the composite and the other variables in the model.  

\textbf{\begin{figure}[t]
    \centering
    \begin{subfigure}[t]{0.45\textwidth}
        \centering
        \includegraphics[width=\textwidth]{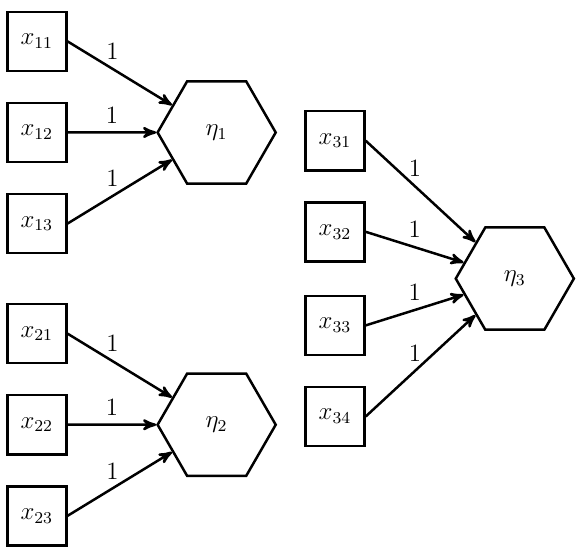}
        \caption{Step 1: Create composite scores as sums}
        \label{fig:twostep1}
    \end{subfigure}
    \hfill
    \begin{subfigure}[t]{0.45\textwidth}
        \centering
        \includegraphics[width=\textwidth]{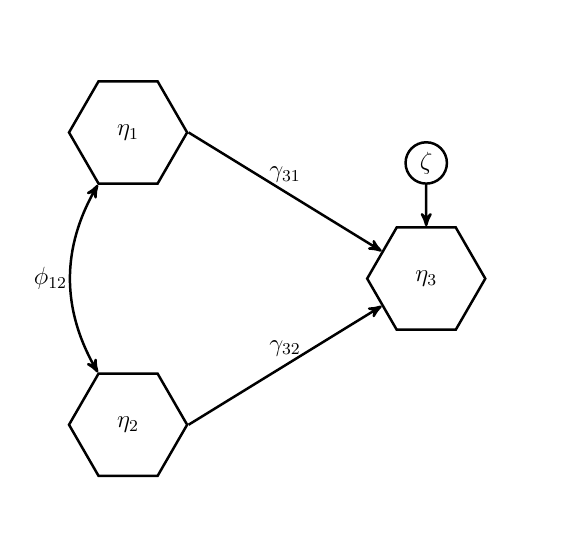}
        \caption{Step 2: Use composite scores in SEM}
        \label{fig:twostep2}
    \end{subfigure}
    \caption{Two-step approach with unit weights applied to the hypothetical scenario}
    \label{fig:twostep}
\end{figure}}

The two-step approach is known for its ease of use. 
It can incorporate composites as predictor and outcome variables.
In the hypothetical scenario described above and as shown in Figure \ref{fig:twostep1}, the composite scores of the predictor composites (i.e., $\eta_{1}$ and $\eta_{2}$) and the outcome composite (i.e., $\eta_{3}$) are created in the first step as the sum of their components.
In the second step, these composite scores are used in SEM to study the relations between the composites, as shown in Figure \ref{fig:twostep2}.
However, given that the composite is created in a separate step rather than modeled within SEM, there are some drawbacks. 
For instance, this approach is limited to studying fixed-weight composites; 
it cannot be used to study unknown-weight composites.
In other words, it is not possible to study composites whose weights are freely estimated. 
Additionally, it is not possible to study composites composed of latent variables as is for example done in the case of Type II second-order constructs \citep{VanRiel2017}.
Moreover, this approach prevents researchers from using the direct maximum-likelihood approach \citep{allison_2003} to address missing values in the components that make up the composites.
Furthermore, excluding the original variables that form the composites can obscure model misspecifications
\citep{mccormick_borgeest_kievit_2022,Schuberth2025,Yu2025}.
This also prevents researchers from testing whether a composite fully transmits the effects of or on its components \citep{Schuberth2025}, which implies that a composite accounts for covariances between its components and other variables in the model.
The latter, for example, is important for testing the synthesis theory \citep{Henseler2021}.

\subsection*{One-step approach} 

Unlike the two-step approach, the \textit{one-step approach} allows researchers to specify composites and their components within SEM \citep{grace_bollen_2008}. 
In this approach, a composite is modeled as a formatively measured latent variable.
Specifically, the components serve as causal indicators of the latent variable.
Thus, their effects on the latent variable constitute the weights. 
Additionally, the variance of the latent variable's error term is fixed to zero, ensuring that the composite is fully composed of its components. 
As a result, the one-step approach can be used to study both fixed-weight and unknown-weight composites. 

The one-step approach models composites directly in SEM and therefore offers advantages in missing data handling and model assessment.
However, its flexibility in modeling composites is limited compared to modeling latent and observed variables in SEM. 
While the one-step approach can specify a composite is a predictor variable, it encounters limitations when a composite is an outcome variable, i.e., when it is affected by other variables in the model, in addition to its components \citep[see, e.g.,][]{Schuberth2023HO}. 
Similarly, using the one-step approach to specify covariances between a composite and other variables in the model is generally very limited \citep{MacCallum1993}.  
As a result, this approach has limitations when testing whether composites fully account for the covariances between their components and other variables in the model, especially when there are two or more predictor composites. 
Additionally, the one-step approach is limited when relaxing the assumption that a composite fully transmits the effects of its components. 

\begin{figure}[t]
    \centering
    \includegraphics[scale = 0.75]{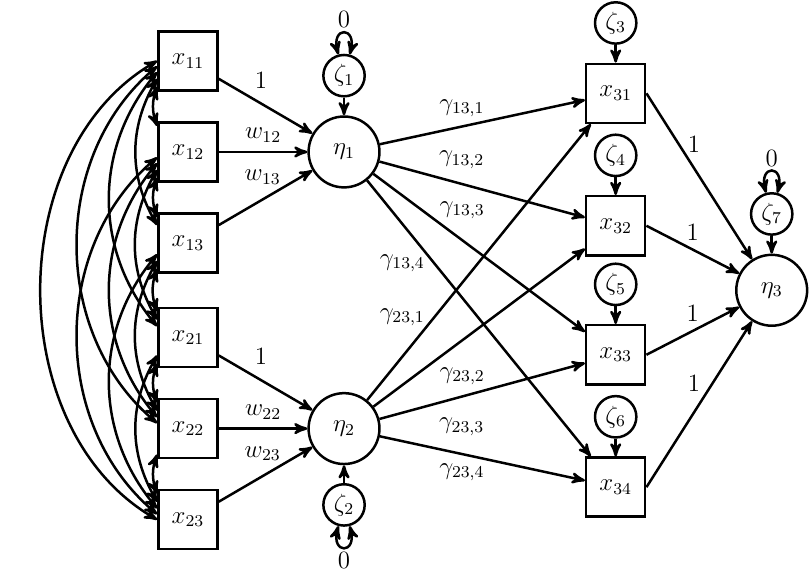}
    \caption{One-step approach applied to the hypothetical scenario}
    \label{fig:onestep}
\end{figure}

Figure \ref{fig:onestep} shows how the one-step approach can be used in the hypothetical scenario.
The composites $\eta_{1}$ and $\eta_{2}$, which predict the composite $\eta_3$, can be modeled using this approach. 
However, there are different views on how to handle covariances between components belonging to different composites. 
For example, \citet{MacCallum1993} fix these covariances to zero, while \citet{grace_bollen_2008} specify them as free model parameters.
For the hypothetical scenario, we adopt the latter approach, allowing for covariances between the components of $\eta_1$ and $\eta_2$.
Additionally, because the weights of these composites are freely estimated, we must fix their scale. 
To this end, we fixed the first weight of these composites to one.
Moreover, the one-step approach has limitations when modeling the composite $\eta_3$ because it cannot distinguish between the components of a composite and its predictors, in our case $x_{31}$ to $x_{34}$ and $\eta_1$ and $\eta_2$. 
Modeling composite $\eta_{3}$ in the same way as composites $\eta_1$ and $\eta_2$  would imply that composite $\eta_{3}$ is composed of its components $x_{31}$ to $x_{34}$ and the two predictor composites $\eta_{1}$ and $\eta_{2}$. 
Clearly, this is not the intended relation that the researcher wants to model. 

An alternative approach is to model the effects of the predictors of a composite on its components rather than modeling them directly on the outcome composite \citep{Schamberger2025}. 
In our scenario, we specified the effects of the two predictor composites $\eta_1$ and $\eta_2$ on the components of $\eta_3$, as illustrated in Figure \ref{fig:onestep}.
This way of including composites is not only very generous in terms of model parameters, but it also does not allow for modeling unknown-weight composites if the composite has no further consequences in the model.
Therefore, in the hypothetical example, we model the composite $\eta_3$ as the sum of its components.
The effects of the composites $\eta_1$ and $\eta_2$ on the composite $\eta_3$ can be derived as indirect effects.

\subsection*{MIMIC approach} 

\citet{Bagozzi_Fornell_Larcker_1981} proposed using a type of MIMIC model to perform a canonical correlation analysis \citep[e.g.,][]{Thompson1984} in SEM and thus study the correlation between two composites. 
In the MIMIC model, a latent variable has both formative and reflective measures \citep{Joreskog1975}. 
In their model, \citet[see their Figure 2]{Bagozzi_Fornell_Larcker_1981} propose modeling the components of the first composite as freely correlated single-indicator variables affecting the latent variable.
Additionally, the components of the second composite serve as reflective measures of the latent variable. 
Notably, the resulting unique terms are allowed to be correlated. 
Finally, the variance of the latent variable's error term is fixed to zero.

The \textit{MIMIC approach} addresses the limitations of the two-step approach by including composite components in the model.
However, there are at least two problems with specifying composites in this way. 
First, the MIMIC approach specifies more parameters than necessary, resulting in fewer degrees of freedom (i.e., overparameterization). 
Second, this specification only explicitly specifies one of the two composites in the model. 
The second composite is expressed through cross-loadings of its components on the first composite. 
Consequently, it is unclear how to flexibly model composites using this approach \citep{Schuberth2023HO}. 
Similarly, it remains unclear how to apply the MIMIC approach to the hypothetical scenario.
Therefore, we must refrain from presenting it graphically.

\subsection*{Composite factor model specification}

\begin{figure}[tb]
    \centering
    \includegraphics[scale = 0.75]{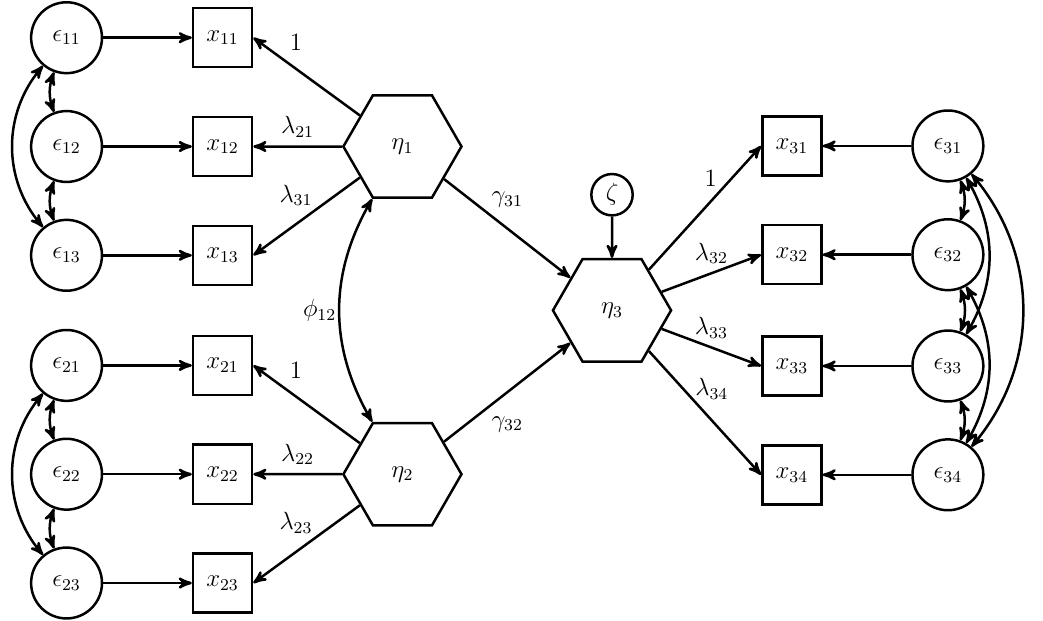}
    \caption{Composite factor model specification applied to the hypothetical scenario}
    \label{fig:compositefactor}
\end{figure}

Another potential way to study composites in SEM is the \textit{composite factor model specification} \citep{Henseler2014}.
This model is equivalent to a common factor model specification with freely correlated unique factors. 
These unique factors capture information that the composite leaves unexplained. 
This is necessary because composites usually do not impose constraints on the covariances among their components. 
Figure \ref{fig:compositefactor} shows the composite factor model specification for our hypothetical scenario. 

The composite factor model can be useful for generating data from populations that follow a composite model \citep[e.g.,][]{Schuberth2018,Schuberth2018a}, and it has theoretical potential for flexible composite modeling.
However, in its current form, the composite factor model specification is not identified \citep{mcintosh_edwards_antonakis_2014}, and therefore  has very limited value for empirical researchers. 
One way to ensure the identification of the free parameters in the composite factor model would be to account for the fact that composites are linear combinations of their components in the model, for example by imposing constraints. 
This is done in other approaches, such as generalized structured component analysis \citep[GSCA,][]{Hwang2004} and partial least squares path modeling \citep[PLS-PM,][]{Wold1982c}, but, to the best of our knowledge, it is not possible in SEM.
Furthermore, this model can be useful for deriving the model-implied variance-covariance matrix and degrees of freedom in the case of PLS-PM, GSCA, and related methods.
The number of degrees of freedom can be determined by subtracting the number of free parameters in the composite factor model and the number of composites from the number of non-redundant sample variances and covariances.

\subsection*{Pseudo-indicator approach} 

\begin{figure}[tb]
    \centering
    \includegraphics[width=0.4\linewidth]{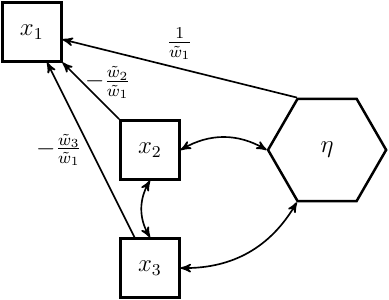}
    \caption{Idea of the pseudo-indicator approach to model a composite}
    \label{fig:ideapseudo}
\end{figure}

The \textit{pseudo-indicator approach} is a flexible approach for modeling both predictor and outcome composites in SEM  \citep{Rose_Wagner_Mayer_Nagengast_2019}. 
This approach exploits the fact that a component can be expressed by means of the composite it forms. 
Specifically, one component, i.e., the pseudo indicator, is expressed as the difference between the composite and the remaining components, as shown in the following equation:
\begin{flalign}
\eta &= \tilde{w}_i x_i + \tilde{\bm w}_{(-i)}'\bm x_{(-i)}\\
    \Leftrightarrow x_i &= \frac{1}{\tilde{w}_i} \eta-\frac{1}{\tilde{w}_i}  \tilde{\bm w}_{(-i)}'\bm x_{(-i)},
\end{flalign}
where $x_i$ is the component selected as pseudo indicator, the vector $\bm x_{(-i)}$ contains the remaining components, and $\tilde{ w}_{i}$ and $\tilde{\bm w}_{(-i)}$ represent the predetermined weights used to form the composite.
In SEM, this idea is modeled by having the pseudo indicator load on a latent variable that represents the composite.
The corresponding loading equals $1/\tilde{w}_i$.
Additionally, the effects of the remaining components on the pseudo indicator must be specified.
Each effect is equal to the negative ratio of the corresponding component's weight to the weight of the pseudo indicator: $-\tilde{\bm w}_{(-i)}/\tilde{w}_i$. 
Finally, the covariances between the remaining components and the composite must be specified.
Figure \ref{fig:ideapseudo} illustrates this idea for a composite with three components, where $x_1$ serves as the pseudo indicator.

The pseudo-indicator approach allows researchers to flexibly model composites in SEM. 
However, it is limited to modeling fixed-weight composites and does not allow researchers to model unknown-weight composites. 
Moreover, the pseudo-indicator approach has been proposed to mimic the results of the two-step approach, specifically the result of the second step. 
Consequently, various covariances between the components and other variables in the model are specified.
Currently, it is unclear how to constrain these covariances to test whether composites fully transmit the effects of or on their components, which limits the applicability of the pseudo-indicator approach \citep{Schuberth2025}.

\begin{figure}[tb]
    \centering
    \includegraphics[scale = 0.75]{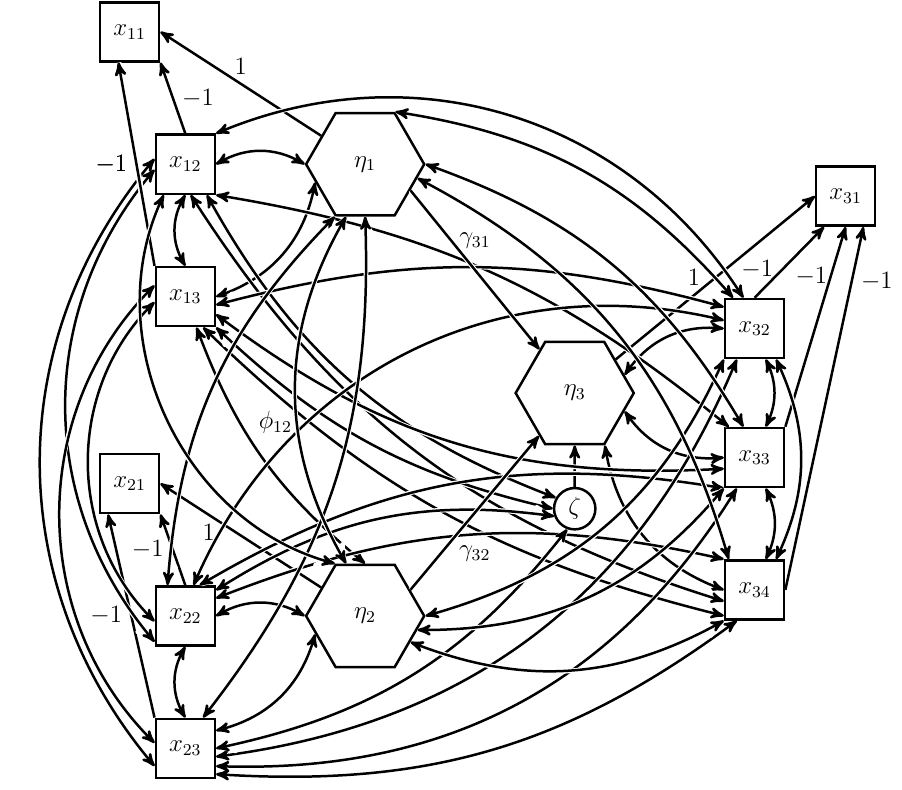}
    \caption{Pseudo-indicator approach with unit weights applied to the hypothetical scenario}
    \label{fig:pseudo}
\end{figure}

Figure \ref{fig:pseudo} illustrates the application of the pseudo-indicator approach to the hypothetical scenario.  
Given that the approach only allows for fixed-weight composites, all composites in this example are specified as sums of their components. 
For the composite $\eta_{1}$, the component $x_{11}$ functions as the pseudo indicator. 
The path coefficients from the components $x_{12}$ and $x_{13}$ to the pseudo-indicator variable $x_{11}$ are fixed to minus one and the path coefficient of the composite on the pseudo indicator is fixed to one. 
Consequently, the pseudo indicator equals the difference between the composite and the remaining components: $x_{11}=\eta_1-x_{12}-x_{13}$.
Furthermore, the covariances between the components, as well as between the components other than the pseudo indicator and the composite $\eta_{1}$ are specified to mimic the second step of the two-step approach.

\subsection*{H–O specification} 

Another way to specify composites in SEM is provided by the \textit{H–O specification} \citep{Schuberth2023HO,henseler_2021}.
It is based on the work of \citet{henseler_2021} and \citet{Ogasawara2007}.
While \citet{henseler_2021} sketched the H–O specification in his book, \citet{Ogasawara2007} showed how to conduct a canonical correlation analysis in SEM by expressing the relationships between the canonical variates and their components in terms of loadings rather than weights.
Therefore, \citet{Schuberth2023HO} introduced it as the H–O specification.

The H–O specification extracts a composite for each component. 
Specifically, it extracts a composite of interest and a set of excrescent variables, as shown in Equation (\ref{eq:HO1}): 
\begin{flalign}
  \bm c=\begin{pmatrix}\eta \\ \bm \nu  \end{pmatrix} = \bm W \bm x, \label{eq:HO1}
\end{flalign}
where the vector $\bm c$ contains the composites, i.e., the composite of interest $\eta$ and the excrescent variables $\bm \nu$.
The square matrix $\bm W$ contains the weights used to form the composites. 
In the H–O specification, the relations between a composite and its components are expressed in terms of composite loadings rather than weights, as shown in Equation (\ref{eq:HO2}):
\begin{flalign}
  \bm W^{-1}\begin{pmatrix} \eta\\ \bm \nu \end{pmatrix} = \bm \Lambda \begin{pmatrix} \eta\\ \bm \nu \end{pmatrix} = \bm x \label{eq:HO2}
\end{flalign}
where $\bm \Lambda$ represents the composite loading matrix. 

To ensure identification of the model parameters, \citet{Schuberth2023HO} proposed fixing the loadings of the excrescent variables to zero in a cascading fashion.
Consequently, all components load on the first excrescent variable while only two components load on the last excrescent variable. 
In addition, one loading per composite is fixed to one to determine the scale of the composites, i.e., the scales of the composite of interest and the excrescent variables.  
Furthermore, the excrescent variables are uncorrelated with each other and with the composite of interest.  
As a result, the composite of interest and the excrescent variables span the space of the components.
Finally, if the weights are freely estimated, the composite of interest must be connected to at least one other variable in the model, in addition to its components; it cannot be isolated.

The H–O specification allows for modeling of unknown-weight composites as predictor and outcome variables, overcoming the limitations of existing approaches.
It can also be used to test whether composites fully transmit the effects of or on their components.
However, it remains unclear how to model fixed-weight composites.
In particular, it is unclear how to constrain the loading matrix to fix the weights of the composite of interest to specific values. 

\begin{figure}[t]
    \centering
    \includegraphics[scale = 0.75]{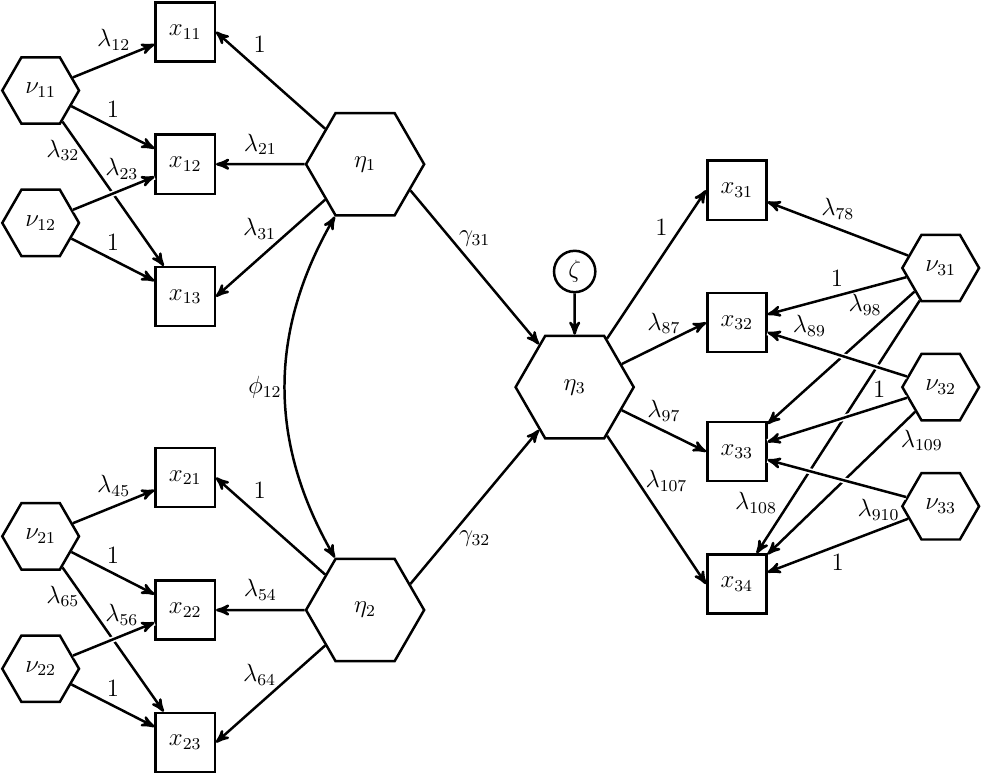}
    \caption{H–O specification with free weights applied to the hypothetical scenario}
    \label{fig:originalHO}
\end{figure}

To further illustrate this approach, we revisit the hypothetical scenario. 
Considering the first set of components $\bm x_1$ in Figure \ref{fig:originalHO}, the composite of interest $\eta_{1}$ and the two excrescent variables $\nu_{11}$ and $\nu_{12}$ collectively span the entire space of the components $x_{11}$ to $x_{13}$. 
The composite loading matrix of the first set of components is shown in Equation (\ref{eq:H03}).
\renewcommand{\kbldelim}{(}
\renewcommand{\kbrdelim}{)}
\begin{flalign}
 \bm \Lambda =  \kbordermatrix{
          & \eta_1        & \nu_{11}      &\nu_{12}     \\ 
    x_{11} &1            &\lambda_{12}  &0\\
    x_{12} &\lambda_{21} &1  &\lambda_{23}   \\
    x_{13} &\lambda_{31} &\lambda_{32}  &1            \\
	} \label{eq:H03}
\end{flalign}
To obtain the final weight estimates, this loading matrix must be inverted; see Equation (\ref{eq:HO2}).
Due to the non-linear relation between the loadings and the weights, it is not possible to find the weights in this model specification intuitively. 

\subsection*{Refined H–O specification} 

The \textit{refined H–O specification} is similar to the H–O specification, but it facilitates model specification and graphical representation \citep{yu_schuberth_henseler_2023}. 
It modifies the composite loading pattern of the excrescent variables.
In contrast to the H–O specification, in the refined H–O specification each excrescent variable is associated with only two components. 
Additionally, the excrescent variables belonging to the same set of components are permitted to be freely correlated.

The refinement of the H–O specification makes it easier to use. 
However, as with the original H–O specification, weights are still not intuitively or conveniently visible in this approach, including in the corresponding software outputs. 
Researchers must perform calculations to obtain estimated weights, which requires some effort. 
Additionally, this approach has limitations when studying fixed-weight composites.
Although weights can be fixed to equal values, it is unclear how to constrain the parameters to fix the weights to specific values \citep{Schuberth2025}.

\begin{figure}[t]
    \centering
    \includegraphics[scale = 0.75]{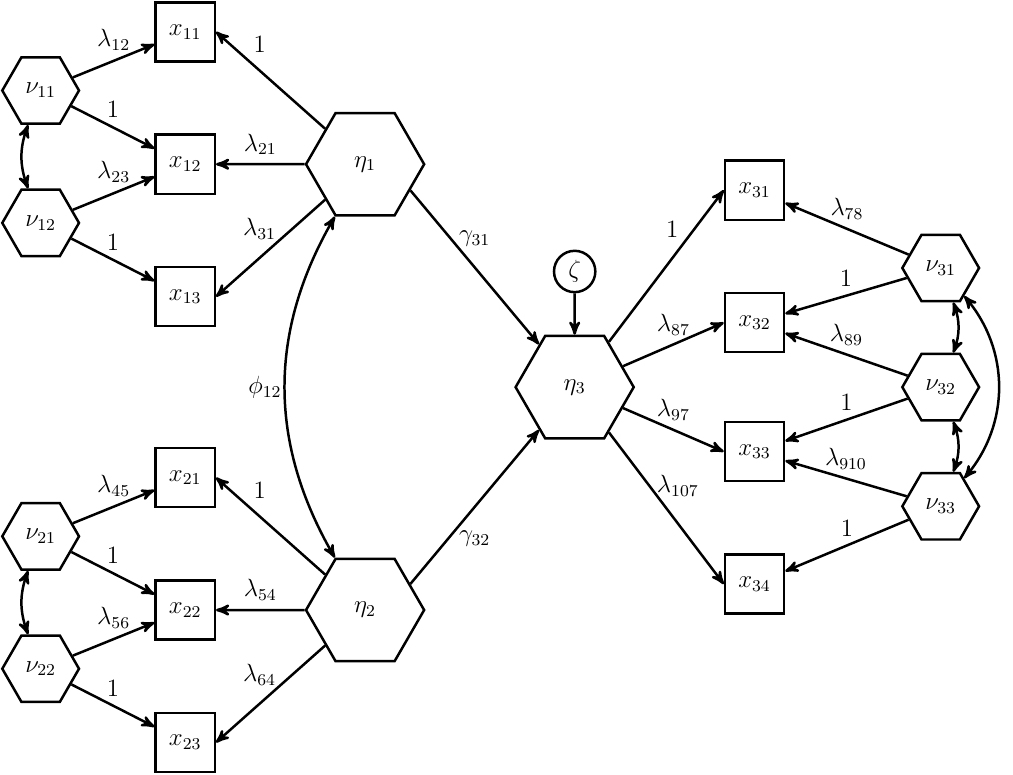}
    \caption{Refined H–O specification with free weights applied to the hypothetical scenario}
    \label{fig:refinedHO}
\end{figure}

Figure \ref{fig:refinedHO} shows the refined H–O specification with free weights for the hypothetical scenario. 
Focusing on the first set of components, the components $x_{11}$ and $x_{12}$ load on the excrescent variable $\nu_{11}$, while the components $x_{12}$ and $x_{13}$ load on the excrescent variable $\nu_{12}$.
Consequently, the composite loading matrix of the original H–O specification, displayed in Equation (\ref{eq:H03}) is modified to the matrix shown in Equation (\ref{eq:H04}). 
Furthermore, the two excrescent variables are allowed to be freely correlated.
\renewcommand{\kbldelim}{(}
\renewcommand{\kbrdelim}{)}
\begin{flalign}
 \bm \Lambda =  \kbordermatrix{
          & \eta_1        & \nu_{11}      &\nu_{12}     \\ 
    x_{11} &1            &\lambda_{12}  &0\\
    x_{12} &\lambda_{21} &1  &\lambda_{23}   \\
    x_{13} &\lambda_{31} &0  &1            \\
	} \label{eq:H04}
\end{flalign}

\begin{figure}[tb]
    \centering
    \includegraphics[scale=0.9]{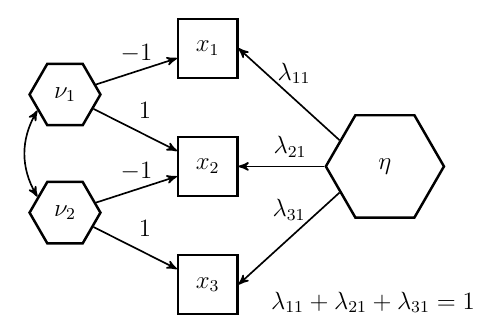}
    \caption{Refined H–O specification to model a composite as the sum of its components}
    \label{fig:refinedsumscore}
\end{figure}

Recently, it has been demonstrated how to fix the weights of the composite of interest to equal values \citep{Schuberth2025}.
For this purpose, the loadings per excrescent variable must be fixed that their sum is zero.
Additionally, the sum of the loadings of the composite of interest must be constrained. 
This constraint determines the weights used to form the composite of interest: $w=1/\sum \lambda_i$.
For example, if the sum is constrained to 1, the composite of interest will be the sum of its components. 
Similarly, constraining the sum of the loadings to the number of components results in a composite of interest that is the average of its components. 
Figure \ref{fig:refinedsumscore} illustrates the refined H–O specification in which the composite of interest is modeled as the sum of three components.

\section*{Enhancements to the H–O specification and its refinement} 
\label{sec:new}

Although the H–O specification and its refined version overcome the limitations of the other approaches, researchers must still calculate the weights of unknown-weight composites manually because these approaches do not intuitively or conveniently display weights. 
Similarly, the two H–O specifications currently have limitations in fixing weights to specific values.
In this section, we address these limitations by enhancing the H–O specification, resulting in two novel specifications. 
First, we propose the phantom-variable H–O specification, which extends the refined H–O specification by adding a phantom variable for each component of a composite. 
Second, we combine the H–O specification and the pseudo-indicator approach to create the blended H–O specification. 

\subsection*{Phantom-variable H–O specification} 

The \textit{phantom-variable H–O specification} starts from the refined H–O specification with unit weights.
Specifically, we apply the constraints presented in the previous section
to ensure that the weights used to form the composite of interest are all equal to one.
This involves fixing the loadings per excrescent variable so that they sum to zero, and constraining the loadings of the composite of interest so that they sum to one.  
Rather than using the original components to build the composite of interest, each component is replaced by a phantom variable $p_i$, which is a latent variable with the original component as its sole indicator.
The variances of the resulting unique terms are fixed to zero. 

Equation (\ref{eq:phantom}) shows the relation between the components $\bm x$ and their phantom variables $\bm p$.
\begin{flalign}
    \bm x = \bm L \bm p \Leftrightarrow \bm L^{-1} \bm x = \bm p \label{eq:phantom}
\end{flalign}
where $\bm L$ is a diagonal matrix containing the loadings of the phantom variables on the components. 
As a result, the relation between the composite of interest and its components is as follows: 
\begin{flalign}
    \eta=\bm 1'\bm p = \bm 1' \bm L^{-1} \bm x = \bm l'\bm x
\end{flalign}
where $\bm l$ is the vector that contains the diagonal elements of $\bm L^{-1}$, i.e., the inverted loadings of the phantom variables on the original components. 
In other words, the weights of the components $\bm x$ to form the composite of interest $\eta$ equal the inverted loadings of the phantom variables $\bm p$ on the components $\bm x$.
Thus, the weights in the phantom-variable H–O specification are obtained more intuitively.
However, this approach has limitations when the weights are close to zero.  
In this case, the respective composite loadings would be quite large, potentially approaching infinity. 

\begin{figure}[tb]
    \centering
    \includegraphics[scale = 0.6]{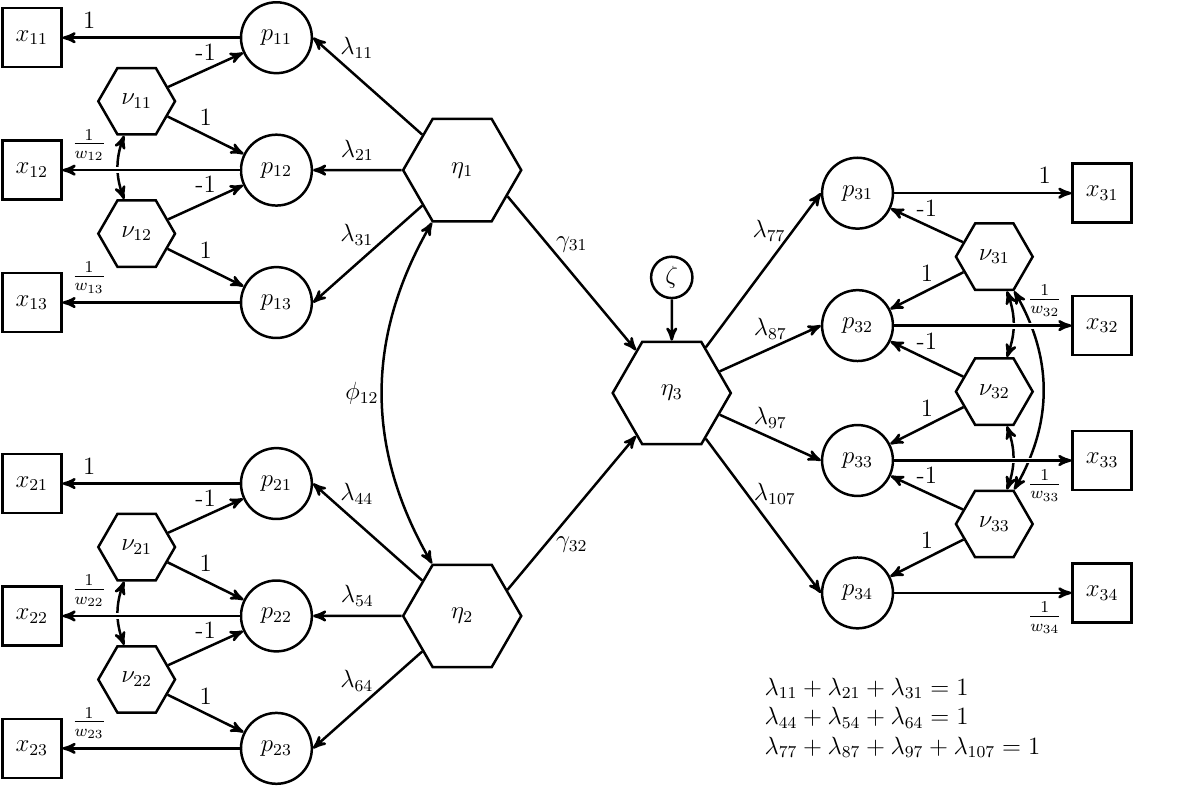}
    \caption{The phantom-variable H–O specification with free weights applied to the hypothetical example}
    \label{fig:phantom}
\end{figure}

The other rules for ensuring model identification are the same as those for the refined H–O specification. 
First, if the weights are freely estimated, the composite of interest cannot be isolated in the structural model; it must be connected to at least one variable within the structural model. 
Additionally, the scale of the composite of interest must be fixed, e.g., by fixing one of the phantom variable loadings, and thus the weight of the corresponding component, to one. 
Second, assuming that the composite of interest accounts for all covariances between its components and the other variables in the model, the excrescent variables are constrained to be uncorrelated with the other exogenous variables in the model.
However. the excrescent variables belonging to the same set of components are allowed to be correlated with each other.
This assumption may be relaxed in certain situations \citep[see, e.g.,][]{grace_bollen_2008,Rose_Wagner_Mayer_Nagengast_2019}.
For instance, if the components are not assumed to have conceptual unity \citep{bollen_noble_2011}, the excrescent variables do not necessarily need to be uncorrelated with other variables in the model.
Similarly, if the composite of interest is not expected to transmit all the collective effects of or on its components, this assumption can be relaxed \citep{Schuberth2025}.
Figure \ref{fig:phantom} shows the phantom-variable H–O specification with free weights for the hypothetical scenario, assuming that the composites of interest account for all covariances between their components. 

\subsection*{Blended H–O specification} 

The \textit{blended H–O specification} blends the (refined) H–O specification and the pseudo-indicator approach.
This allows researchers to directly obtain the weights used to form the composite of interest. 
Like the (refined) H–O specification, this approach extracts as many composites as there are components, i.e., one composite of interest and a set of excrescent variables.
Additionally, it borrows the idea of the pseudo-indicator approach to express one of the components, i.e., the pseudo indicator, as the difference between the composite of interest and the remaining components, as shown in the following equation:
\begin{flalign}
\eta &= w_i x_i + \bm w_{(-i)}' \bm x_{(-i)}\\
\Leftrightarrow
x_i &= \frac{1}{w_i}\eta - \frac{1}{w_i}\bm w_{(-i)}'\bm x_{(-i)},\label{eq:blended}
\end{flalign}
where $x_i$ is the component chosen as the pseudo indicator, the vector $\bm w_{(-i)}$ contains all weights used to form the composite of interest except for the weight of the chosen pseudo indicator $w_i$, and the vector $\bm x_{(-i)}$ contains all components except the pseudo indicator.  

\begin{figure}[tb]
    \centering
    \includegraphics[width=0.4\linewidth]{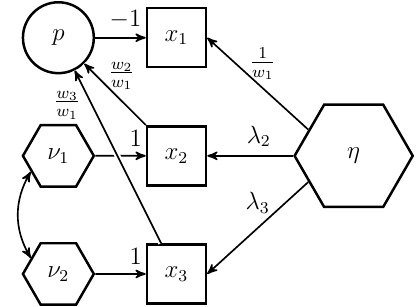}
    \caption{Idea of the blended H–O specification to model a composite}
    \label{fig:ideablended}
\end{figure}

Figure \ref{fig:ideablended} further illustrates this idea.
In contrast to the refined H–O specification, in which each excrescent variable loads on two components, in the blended H–O specification each excrescent variable (in our case $\nu_1$ and $\nu_2$) loads on only one component, except for the component that serves as the pseudo indicator ($x_1$ in our case), which is not connected to an excrescent variable. 
Additionally, a phantom variable $p$ is created to reverse the sign of the effects of the other components on the pseudo indicator.
Consequently, the effects of the other components on the phantom variable equal the weight ratios. 
Therefore, when the weight of the pseudo indicator, i.e., the loading of the composite of interest on the pseudo indicator, is fixed to one, the remaining weights are directly visible in the model specification as the effects of the components on the phantom variable.
Otherwise, these effects represent the ratio of the corresponding weight to the weight of the pseudo indicator.

\begin{figure}[tb]
    \centering
    \includegraphics[scale = 0.75]{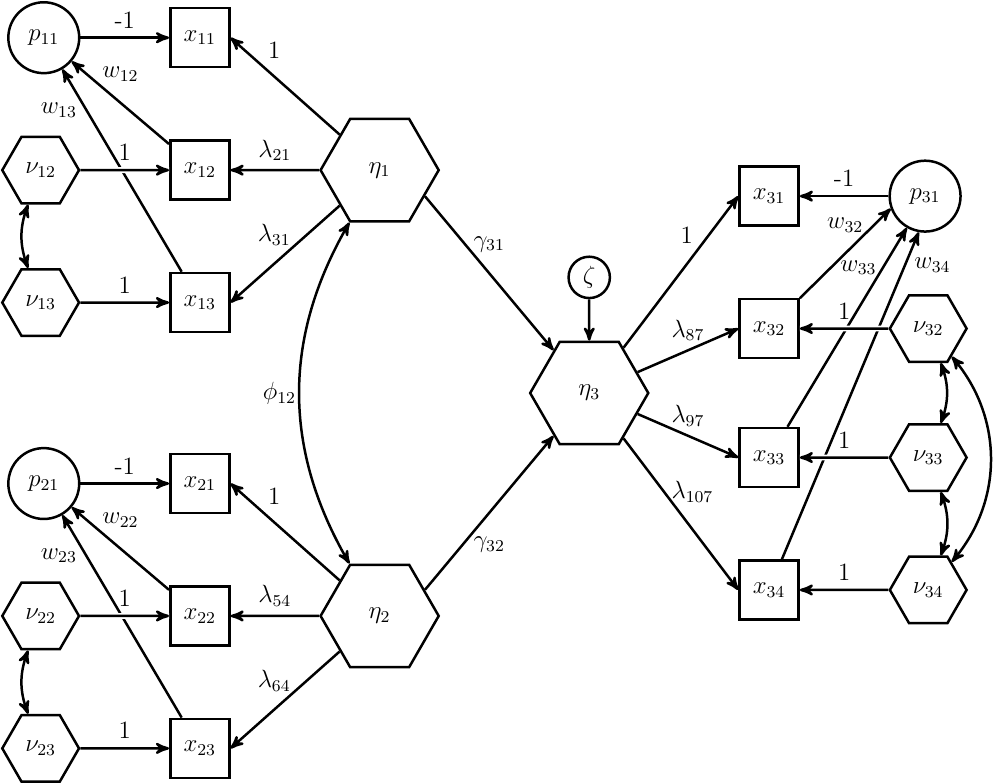}
    \caption{The blended H–O specification with free weights applied to the hypothetical scenario}
    \label{fig:blended}
\end{figure}

The remaining rules for model identification are the same as those of the refined H–O specification: 
First, if the weights are freely estimated, the composite of interest cannot be isolated; it must be connected to other model variables, in addition to its components, in the model.
Second, if the composite of interest accounts for the covariances between its components and other model variables, the excrescent variables must be uncorrelated with the model's other exogenous variables. 
Figure \ref{fig:blended} shows this specification for the hypothetical scenario, assuming the composites of interest are unknown-weight composites that account for all covariances between their components and other model variables. 

\section*{Illustrative example} \label{sec:example}

\subsection*{Description}

To demonstrate and compare the phantom-variable and the blended H–O specifications with existing approaches, we present an illustrative example.
Specifically, we use the empirical study and dataset of \citet{burgoyne_hambrick_altmann_2019}.
This dataset is publicly available\footnote{The dataset can be accessed via the Open Science Framework: \url{https://
osf.io/ndquz/?view_only=87cd1ad87c9e4fabba3e
21c609ce9a14}} and originally consists of 248 participants.
To better compare the different approaches, we excluded four rows with missing values, resulting in 244 complete observations of the following variables: \textit{Letter Sets (Lett)}, which is the number of correct responses to a task in which participants identify the letter set that differs from others based on a given pattern; \textit{Raven’s Progressive Matrices (Rave)}, which is the number of correct responses to a pattern-completion task involving visual matrices; \textit{Operation Span (Ospa)}, which is the number of letters that were correctly recalled while solving interleaved math problems; \textit{Symmetry Span (Sspa)}, which is the number of square locations that were correctly recalled after making symmetry judgments on visual patterns;
 \textit{UNRAVEL error rate (UE)}, which is the proportion of incorrect rule applications in the UNRAVEL task, \textit{UNRAVEL response time (UR)}, which is the average response time for correct trials in the UNRAVEL task; \textit{Letter Wheel error rate (LE)}, which is the proportion of incorrect alphabetizations in the Letter Wheel task; and \textit{Letter Wheel response time (LR)}, which is the average response time for correct trials in the Letter Wheel task. For a detailed explanation of the variables, we refer the reader to the original study of \citet{burgoyne_hambrick_altmann_2019}.


In their study, \citet{burgoyne_hambrick_altmann_2019} employed hierarchical regression analysis and SEM to examine the relations between \textit{working memory capacity (WMC)}, \textit{fluid intelligence (Gf)}, \textit{placekeeping error rate (PE)}, and \textit{placekeeping response time (PR)}.
In our illustrative example, we focus on their hierarchical regression analysis, for which composites were formed as the average z-scores of the corresponding measures for \textit{Gf}, \textit{WMC}, \textit{PE}, and \textit{PR}.
In particular, the composite \textit{Gf} is made up of \textit{Lett} and \textit{Rave}, the composite \textit{WMC} is composed of \textit{Ospa} and \textit{Sspa}, the composite \textit{PE} is made up of \textit{UE} and \textit{LE}, and the composite \textit{PR} is composed of \textit{UR} and \textit{LR}. 
Subsequently, they estimated various regression models. 
For our illustration, we focus on their model from Step 2: the regression model in which \textit{WMC}, \textit{GR}, and \textit{PR} predict \textit{Gf}:
\begin{flalign}
    Gf= \gamma_{1}WMC + \gamma_{2}PE + \gamma_{3}PR + \zeta
\end{flalign}


\subsection*{Model specifications}

To replicate their regression model from the second step of the hierarchical regression analysis in SEM, we apply 
\begin{enumerate*}[label=(\arabic*)]
\item the two-step approach with composites modeled as averages;\label{app:twostep}
\item the one-step approach with composites modeled as averages;\label{app:onestep}
\item the pseudo-indicator approach with composites modeled as averages;\label{app:pseudo}
\item the refined H–O specification with composites modeled as averages;\label{app:refinedav}
\item the phantom-variable H–O specification with composites modeled as averages;\label{app:phantomav} and 
\item the blended H–O specification with composites modeled as averages.\label{app:blendedav}
\end{enumerate*}
The original H–O specification was not applied in this case because it is unclear how to assign specific values to the weights using this approach. 
Similarly, the original one-step approach does not allow us to model composites as variables affected by variables other than their components, such as \textit{Gf} in our illustrative example. 
Therefore, we applied the modified one-step approach, presented previously.
Moreover, we specified the Approaches \ref{app:pseudo} to \ref{app:blendedav} in such a way that they mimic the results of the second step of the two-step approach, i.e., not all covariances between the components are accounted for by the composites of interest. 

The original study by \citet{burgoyne_hambrick_altmann_2019} postulated that, by using equal weights, all components contribute equally to the creation of the composites. 
Alternatively, the weights could be estimated to yield an optimal model fit. 
This makes the weighting of the indicators an empirical question rather than a normative statement. 
This allows analysts to reassess whether all components play their intended role in the model.
Therefore, in addition to fixed-weight composites, we estimated the model with unknown-weight composites using 
\begin{enumerate*}[label=\arabic*), start=7]
\item the (refined) H–O specification with free weights;\label{app:refinedfree}
\item the phantom-variable H–O specification with free weights;\label{app:phantomfree} and 
\item the blended H–O specification with free weights.\label{app:blendedfree}
\end{enumerate*}
We did not consider the two-step and pseudo-indicator approaches in our illustrative example because they do not permit studying unknown-weight composites.
Similarly, we cannot use our modified one-step approach to model unknown-weight composites that do not affect any other variable in the model, such as \textit{Gf}.
Therefore, we did not consider this approach either.
Furthermore, the composites in this empirical study are made up of two components. 
In this case, the original and the refined H–O specifications are identical.
Consequently, we only report the results of one of the two approaches.

\subsubsection*{Approach 1): Two-step approach with composites modeled as averages}

\begin{figure}[tb]
    \centering
    \begin{subfigure}[t]{0.45\textwidth}
        \centering
        \includegraphics[width=\textwidth]{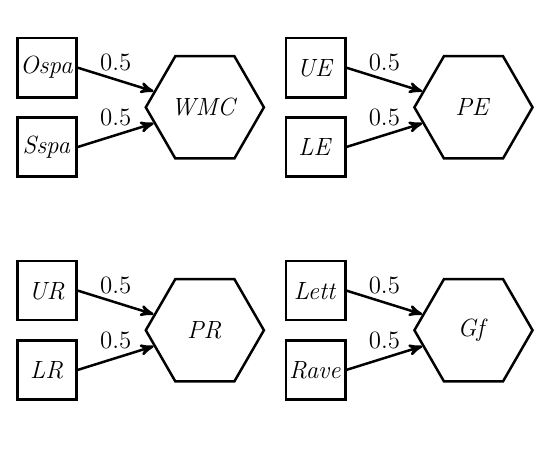}
        \caption{Step 1: Create composite scores as averages}
        \label{fig:modelts_1}
    \end{subfigure}
    \hfill
    \begin{subfigure}[t]{0.45\textwidth}
        \centering
        \includegraphics[width=\textwidth]{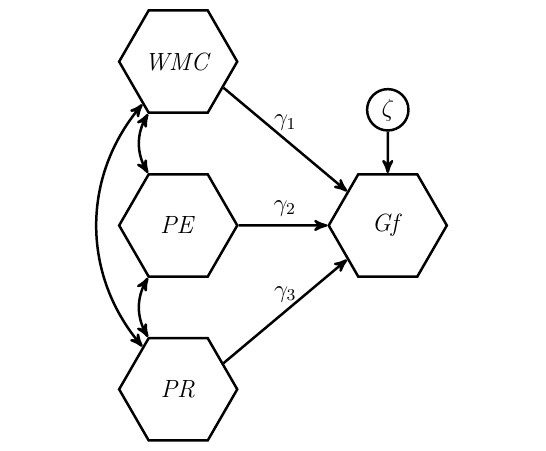}
        \caption{Step 2: Use the composite scores in SEM}
        \label{fig:modelts_2}
    \end{subfigure}
    \caption{Two-step approach with composites modeled as averages}
    \label{fig:modelts}
\end{figure}

The two-step approach mimics the hierarchical regression performed in \citet{burgoyne_hambrick_altmann_2019}.
In the first step, composite scores are created by averaging the standardized variables. 
Specifically, we obtain the composite score for \textit{WMC} by averaging \textit{Ospa} and \textit{Sspa}. 
We obtain the scores for the other composites similarly, by averaging the corresponding components, as shown in Figure \ref{fig:modelts_1}.
 In the second step, we use these composite scores as input for SEM to estimate the relations between the composites. 
 The model from the second step is shown in Figure \ref{fig:modelts_2}. 

\subsubsection*{Approach 2): One-step approach with composites modeled as averages}

\begin{figure}[tb]
    \centering
    \includegraphics[width=0.8\textwidth]{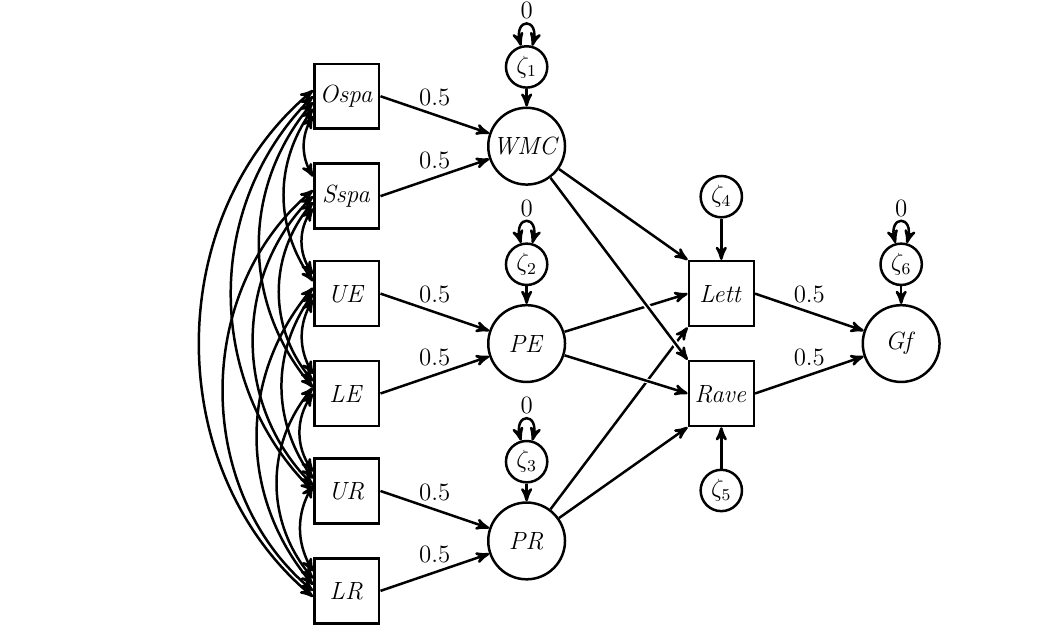}
    \caption{One-step approach with composites modeled as averages}
    \label{fig:modelos}
\end{figure}

As a second approach, we apply the one-step approach shown in Figure \ref{fig:modelos}. 
Specifically, we specify the composites as error-free dependent latent variables, fixing the weights of their respective components to $0.5$ to ensure that the composites are created as the averages of their components. 
Considering the components of the predictor composites, i.e., \textit{WMC}, \textit{PE}, and \textit{PR}, we follow \citet{grace_bollen_2008} and allow them to freely covary.  
Because the one-step approach cannot directly model a composite affected by variables other than its components (\textit{Gf} in our case), we use the modified one-step approach presented previously 
and specify the effects of the predictor composites \textit{WMC}, \textit{PE}, and \textit{PR} on the components of   \textit{Gf}, i.e., \textit{Lett} and \textit{Rave}.

\subsubsection*{Approach 3): Pseudo-indicator approach with composites modeled as averages}

\begin{figure}[tb]
    \centering
    \includegraphics[width=0.8\textwidth = 0.5]{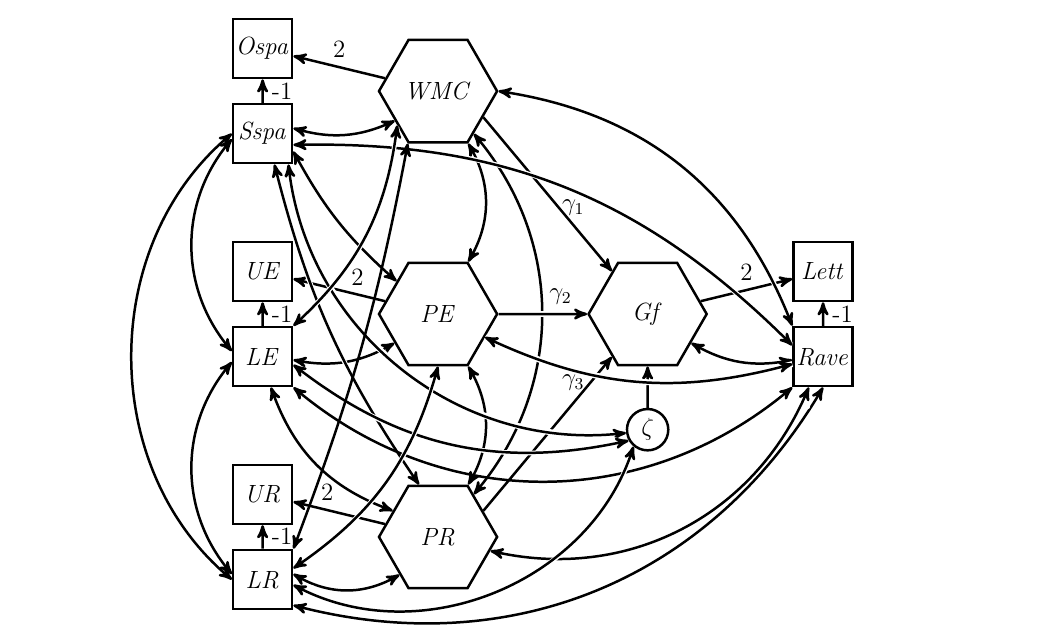}
    \caption{Pseudo-indicator approach with composites modeled as averages}
    \label{fig:modelpimfull}
\end{figure}

As a third approach, we apply the pseudo-indicator approach, as illustrated in Figure \ref{fig:modelpimfull}.
All composites are created as averages of their respective components. 
For example, consider the composite \textit{Gf} and its two components.
We select the component \textit{Lett} as the pseudo indicator.
Moreover, we fix the effect of the composite on the pseudo indicator to two and  fix the effect of the remaining component on the pseudo-indicator to minus one.
As a consequence, \textit{Gf} is formed as the average of its two standardized components.
We do the same for the other three composites. 
Additionally, we must specify several covariances to eliminate unintended constraints imposed by including the composites' components. 
Thus, the pseudo-indicator approach mimics the results of the second step of the two-step approach.

\subsubsection*{Approach 4): Refined H–O specification with composites modeled as averages}

\begin{figure}[tb]
        \centering        \includegraphics[width=0.8\textwidth]{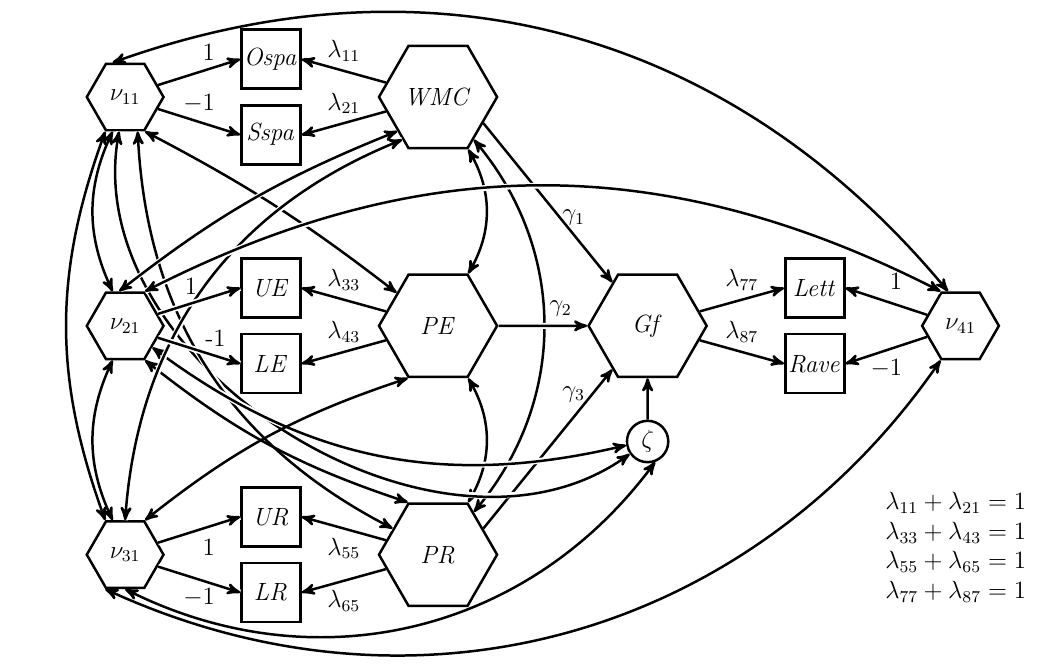}
        \caption{Refined H–O specification with composites as averages}
        \label{fig:ho_u}
    \end{figure}

We further apply the refined H–O specification.
Taking the two components that make up the composite \textit{Gf} as an example, we extract one excrescent variable in addition to the composite of interest \textit{Gf}.
We fix the composite loadings of the excrescent variable on its two components to one and minus one, respectively, and constrain the sum of the loadings of \textit{Gf} on the two components to two. 
In addition, the excrescent variable is uncorrelated with \textit{Gf}, but is allowed to be correlated with the other composites of interest and excrescent variables.
We perform a similar process for the other composites, as illustrated in Figure \ref{fig:ho_u}.

\subsubsection*{Approach 5): Phantom-variable H–O specification with composites modeled as averages}

\begin{figure}[tb]
    \centering
    \includegraphics[width=0.8\textwidth]{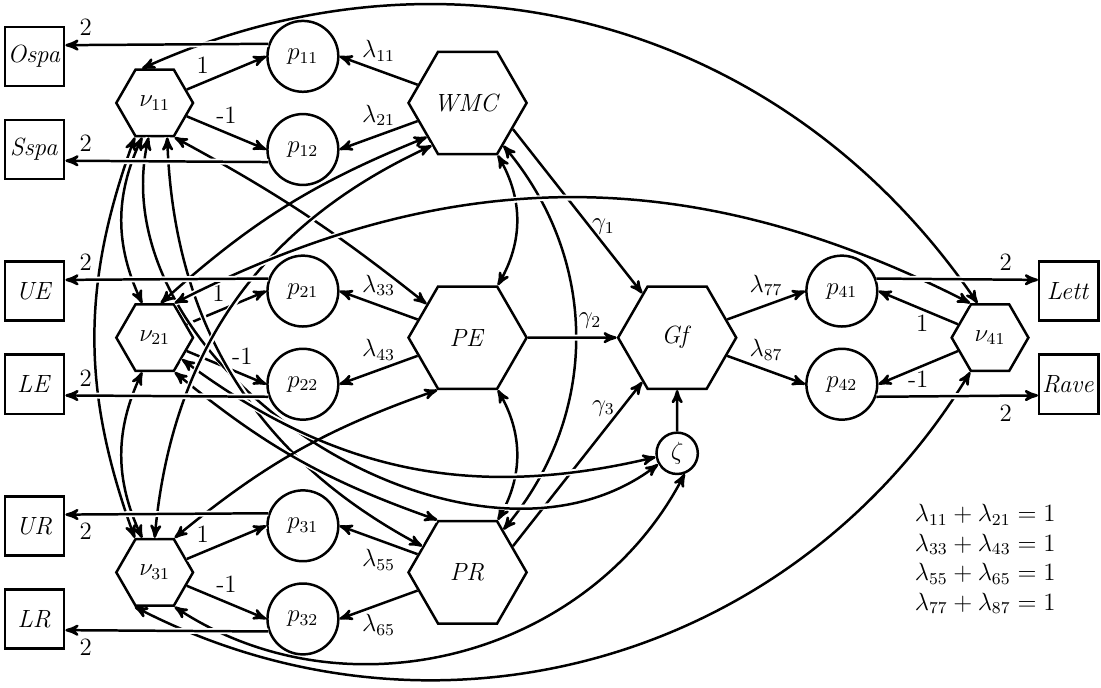}
        \caption{Phantom-variable H–O specification with average weights}
        \label{fig:hopv_u}
\end{figure}

As a fifth approach, we employ the phantom-variable H–O specification.
In contrast to the refined H–O specification, this specification replaces the original components with phantom variables, which are latent variables that use the corresponding component as single indicator. 
As a result, modeling the composites of interest as averages of their corresponding components is straightforward. 
For the composite \textit{Gf}, we first specify \textit{Gf} as the sum of its two phantom variables $p_{41}$ and $p_{42}$. 
To achieve this, we fix the sum of the loadings of \textit{Gf} on the two phantom variables to one. 
Additionally, we fix the two loadings of the excrescent variable on the phantom variables to one and minus one, ensuring that they sum to zero.  
To ensure that \textit{Gf} is the average of the two components, we fix the loadings of the phantom variables on their corresponding indicators to two. 
We proceed similarly for the other composites, as shown in Figure \ref{fig:hopv_u}.
Furthermore, we must specify various covariances between the excrescent variables to relax the assumption that all covariances between the components are accounted for by the composites of interest. 

\subsubsection*{Approach 6): Blended H–O specification with composites modeled as averages}

\begin{figure}[tb]
    \centering       \includegraphics[width=0.8\textwidth]{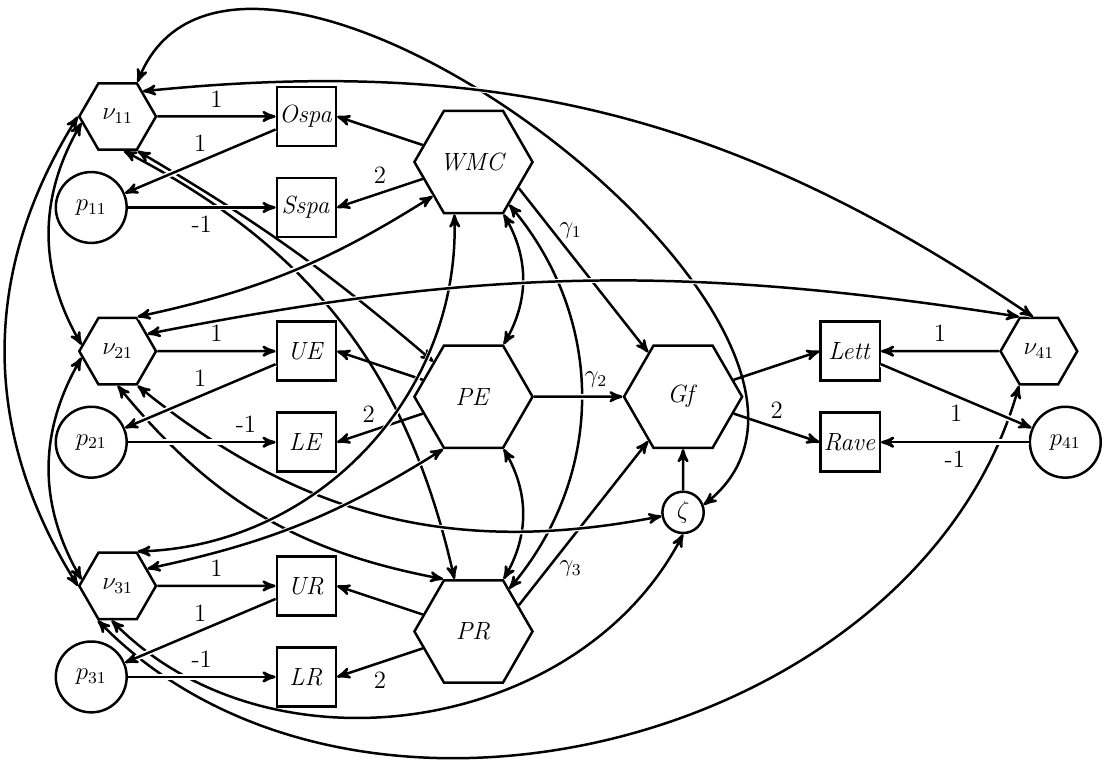}
        \caption{Blended H–O specification with average weights}
        \label{fig:bho_u}
\end{figure}

As a final approach, we use the blended H–O specification to model the composites as averages.
Considering the components of \textit{Gf}, we select \textit{Rave} as the pseudo indicator. 
To create \textit{Gf} as the average of its two components, we fix the loading of \textit{Gf}, which equals one over the weight of \textit{Rave}, to two. 
In addition, we introduce a phantom variable that mediates the effect of \textit{Lett} on \textit{Rave}.
We fix the effect of \textit{Lett} on the phantom variable to one, which equals the weight of \textit{Rave} divided by the weight of \textit{Lett}, meaning the weights of \textit{Lett} and \textit{Rave} are equal.
We proceed similarly for the other composites, as shown in Figure \ref{fig:bho_u}.
Moreover, we specify several covariances of the excrescent variable to relax the assumption that all covariances between components are accounted for by the composites of interest. 

\subsubsection*{Approach 7): (Refined) H–O specification with free weights}

\begin{figure}[tb]
        \centering        \includegraphics[width=0.8\textwidth]{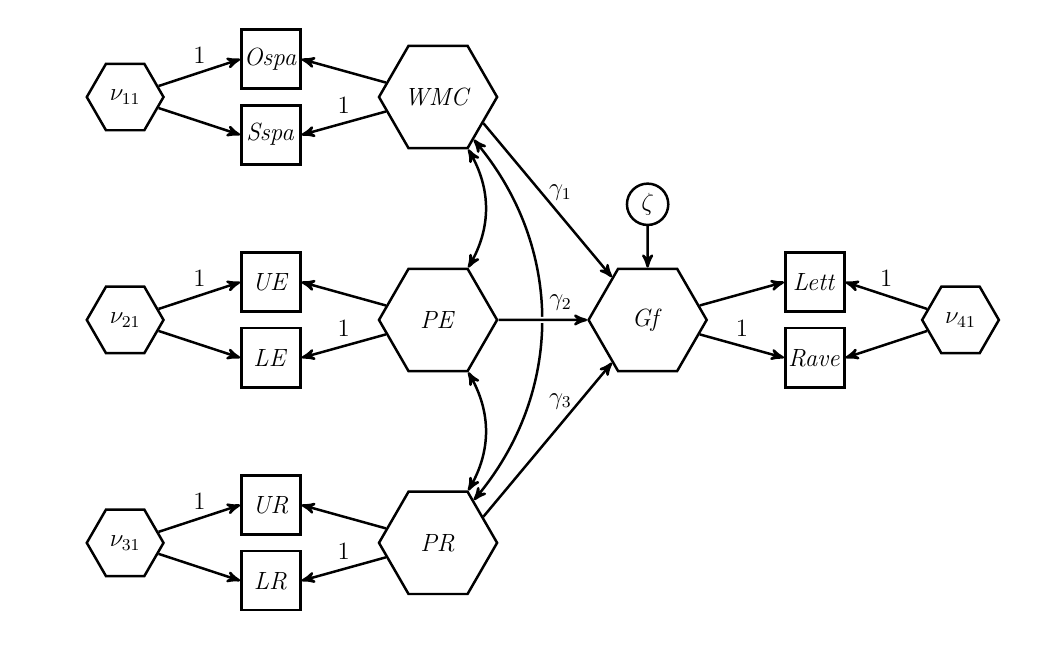}
        \caption{Refined H–O specification with free weights}
        \label{fig:ho_f}
    \end{figure}

The seventh approach is the refined H–O specification with free weights.
Considering the two components of \textit{Gf}, we extract two composites:  \textit{Gf} and an excrescent variable $\nu_{41}$ in our case.
To fix the scale of the two composites, we fix the loadings of \textit{Gf} on \textit{Rave} and of the excrescent variable on \textit{Lett} to one. 
In addition, the excrescent variable is uncorrelated with all the other variables in the model.
Consequently, the composite \textit{Gf} accounts for the covariances between its components and the other variables in the model. 
As shown in Figure \ref{fig:ho_f}, a similar process is performed for the components of the other composites.

\subsubsection*{Approach 8): Phantom-variable H–O specification with free weights}
    
\begin{figure}[H]
    \centering
    \includegraphics[width=0.8\textwidth]{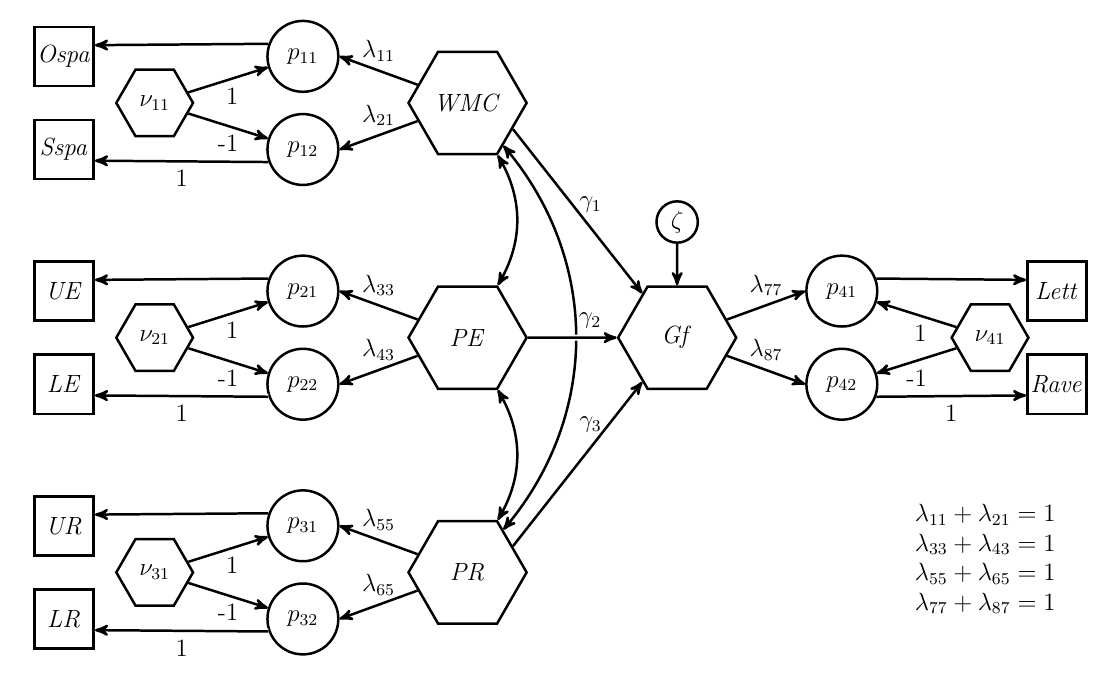}
        \caption{Phantom-variable H–O specification with free weights}
        \label{fig:hopv_f}
\end{figure}

As an eighth approach, we use the phantom-variable H–O specification with free weights. 
Similar to the phantom-variable H–O specification with composites modeled as averages, we fix the loadings of the composites of interest and excrescent variables in a way that creates the composites of interest as the sum of their corresponding phantom variables, as illustrated in Figure \ref{fig:hopv_f}.
Unlike the phantom-variable H–O specification with fixed weights, in this approach, only the loading of one phantom variable on its indicator per composite of interest is fixed to one.
Consequently, the weight of the corresponding component to form the composite of interest is also fixed to one. 
Additionally, the loadings of the remaining phantom variables are freely estimated. 
As discussed in the section on the phantom-variable approach,
these loadings are equal to the corresponding inverted weights.
For the composite of interest \textit{Gf}, the loading of the phantom variable $p_{41}$ on the component \textit{Lett} is equal to $1/w_\textit{Lett}$, while the weight of \textit{Rave} to form \textit{Gf} is fixed to one to determine the scale of \textit{Gf}.
Furthermore, the covariances between the excrescent variables and other model variables are fixed to zero.
Consequently, the composites of interest account for the covariances between their components and the other variables in the model. 

\subsubsection*{Approach 9): Blended H–O specification with free weights}

\begin{figure}[H]
    \centering       \includegraphics[width=0.8\textwidth]{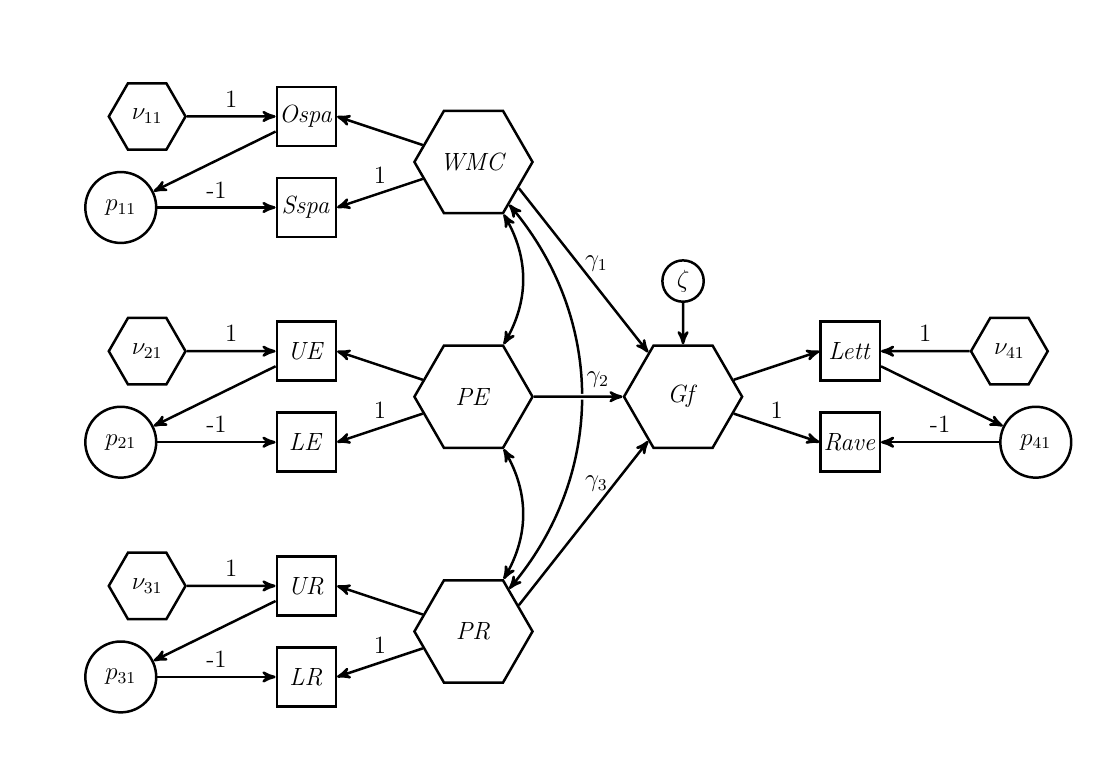}
        \caption{Blended H–O specification with free weights}
        \label{fig:bho_f}
\end{figure}

Finally, we apply the blended H–O specification with free weights.
Considering the components that make up \textit{Gf}, we set the loading of \textit{Gf} on \textit{Rave} to one, thereby setting the weight of \textit{Rave} to form the composite \textit{Gf} to one. 
Consequently, the weight of \textit{Lett} is a free model parameter equal to the effect of \textit{Lett} on the phantom variable $p_{41}$.
Furthermore, we fix the covariances between the excrescent variables and the other variables in the model to zero. 
As a result, the composites of interest account for the covariances between their components and the other model variables. 
The complete specification is shown in Figure \ref{fig:bho_f}.

\subsection*{Results}

All model estimations were performed in the  statistical computing environment R \citep[][Version 4.5.1]{Rsoftware} using the lavaan package \citep{lavaan}.
All estimations terminated successfully. 
The complete R code for all model estimations is available on the Open Science Framework: \url{https://osf.io/tdvwq/?view_only=f870a5b3592e45e8802e9db077ca9283X}.

The results of the approaches used to model the composites as averages of their standardized indicators are reported in Table \ref{tab:resultsunit}.\footnote{Our results for the two-step approach differ slightly from those reported in \citet{burgoyne_hambrick_altmann_2019} because we used listwise deletion to address missing values, whereas they removed missing values when calculating the averages.}
Because each composite is formed by two components, the weights are 0.5 for all approaches. 
Additionally, all approaches except the one-step approach (i.e., the blended H–O specification, the phantom variable H–O specification, the refined H–O specification and the pseudo-indicator approach) can mimic the results of the two-step approach. 
These approaches produce identical path coefficient estimates, including their standard errors (SEs), $\chi^2$ statistic, degrees of freedom (dfs), root mean square error of approximation (RMSEA) and coefficient of determination ($R^2$) for \textit{Gf}.
Additionally, the standardized root mean square residual (SRMR) is equal to zero across the five approaches. 
This is because the model from the second step of the two-step approach is saturated, which is not generally the case.\footnote{To ensure that the pseudo-indicator approach and the H–O specifications produce the same SRMR as the two-step approach, the calculation of the SRMR needs adjustment \citep{Rose_Wagner_Mayer_Nagengast_2019}.}
Furthermore, the model information criteria, such as the Akaike and Bayesian information criteria, differ between the two-step approach and the other four approaches. 
This difference is due to the larger number of observed variables and increased number of model parameters in the four other approaches \citep{Rose_Wagner_Mayer_Nagengast_2019}. 
Unlike the other approaches, the one-step approach does not directly estimate the effects between the composites.
This is because, we cannot specify the effects of the predictor composites on the outcome composite in this approach.
Therefore, we specified the effects of the predictor composites on the components of \textit{Gf}.
However, the effects of \textit{WMC}, \textit{PE}, and \textit{PR} on \textit{Gf} can be derived as indirect effects. 
For example, the effect of \textit{WMC} on \textit{Gf} can be estimated as $0.5\cdot (0.186 + 0.129)$, where the parentheses contain the effects of \textit{WMC} on \textit{Lett} and \textit{Rave}, and 0.5 equals the weight of \textit{Lett} and \textit{Rave}, respectively, to form \textit{Gf}.
As can be seen in Table \ref{tab:resultsunit}, the derived effects equal those between composites produced by the other approaches. 
However, their SEs are different. 
Moreover, the one-step approach shows seven degrees of freedom (dfs) and thus imposes more constraints on the components' variance-covariance matrix than the other approaches. 
This is also indicated by the positive values of the fit statistics such as the $\chi^2$ statistic, RMSEA and SRMR.

\begin{landscape}
\footnotesize
\begin{longtable}{lccccccccccccccccc}
\caption{Results for the illustrative example with composites modeled as averages} \label{tab:resultsunit} \\
\toprule
&\multicolumn{2}{p{2.2cm}}{\centering \pbox{20cm}{\relax\ifvmode\centering\fi Two-step\\ approach}}&& 
\multicolumn{2}{p{2.2cm}}{\centering \pbox{20cm}{\relax\ifvmode\centering\fi Modified one-step\\ approach}}&& 
\multicolumn{2}{p{2.2cm}}{\centering \pbox{20cm}{\relax\ifvmode\centering\fi Pseudo-indicator\\ approach}}&&
\multicolumn{2}{p{2,2cm}}{\centering \pbox{20cm}{\relax\ifvmode\centering\fi Refined H–O\\specification}}&& 
\multicolumn{2}{p{2.2cm}}{\centering \pbox{20cm}{\relax\ifvmode\centering\fi Phantom-variable\\ H–O specification }}&& 
\multicolumn{2}{p{2.2cm}}{\centering \pbox{20cm}{\relax\ifvmode\centering\fi Blended H–O\\specification}}\\
\cline{2-3} \cline{5-6} \cline{8-9} \cline{11-12} \cline{14-15} \cline{17-18}
 & Est & SE && Est & SE && Est & SE && Est & SE && Est & SE && Est & SE\\[-9pt]
 \textbf{Path coefficients}&  &  &&  & &&  &  &&  &  &&  &  &&  &  \\
$\gamma_1$: \textit{WMC} $\rightarrow$ \textit{Gf} & \phantom{-}0.157 & 0.063 && \phantom{-}0.157$^a$ & 0.058$^a$  && \phantom{-}0.157 & 0.063 && \phantom{-}0.157 & 0.063 && \phantom{-}0.157 & 0.063 &&  \phantom{-}0.157 & 0.063\\
$\gamma_2$: \textit{PE} $\rightarrow$ \textit{Gf} & -0.261 & 0.061 && -0.261$^a$ & 0.056$^a$ && -0.261 & 0.061 && -0.261 & 0.061 && -0.261 & 0.061 &&  -0.261 & 0.061\\
$\gamma_3$: \textit{PR} $\rightarrow$ \textit{Gf} & -0.236 & 0.056 && -0.236$^a$ & 0.052$^a$ && -0.236 & 0.056 && -0.236 & 0.056 && -0.236 & 0.056 &&  -0.236 & 0.056 \\[3pt]
\multicolumn{15}{l}{\textbf{Composite covariances}}\\
\textit{WMC} $\leftrightarrow$ \textit{PE} & -0.222 & 0.043 && N.A. & N.A.  && -0.222 & 0.043 && -0.222 & 0.043 &&-0.222 & 0.043 &&-0.222 & 0.043 \\
\textit{WMC} $\leftrightarrow$ \textit{PR} & -0.082 & 0.042 &&  N.A. & N.A.  && -0.082 & 0.042 && -0.082 & 0.042 && -0.082 & 0.042 &&  -0.082 & 0.042 \\
\textit{PE} $\leftrightarrow$ \textit{PR} & -0.006 & 0.043 && N.A. & N.A.  && -0.006 & 0.043 && -0.006 & 0.043 && -0.006 & 0.043 &&  -0.006 & 0.043 \\[3pt]
\multicolumn{15}{l}{\textbf{Fit statistics}}\\
$\chi^2$ statistic ($df$) &  \multicolumn{2}{c}{0.000 (0)}&& \multicolumn{2}{c}{29.135 (7)} && \multicolumn{2}{c}{0.000 (0)}&&  \multicolumn{2}{c}{0.000(0)} &&\multicolumn{2}{c}{0.000 (0)}&& \multicolumn{2}{c}{0.000 (0)}\\
RMSEA & \multicolumn{2}{c}{0.000} && \multicolumn{2}{c}{0.114} && \multicolumn{2}{c}{0.000}&&  \multicolumn{2}{c}{0.000} &&  \multicolumn{2}{c}{0.000}&&\multicolumn{2}{c}{0.000}\\
SRMR & \multicolumn{2}{c}{0.000} && \multicolumn{2}{c}{0.048} &&\multicolumn{2}{c}{0.000}&&  \multicolumn{2}{c}{0.000} &&  \multicolumn{2}{c}{0.000}&&\multicolumn{2}{c}{0.000}\\
AIC &  \multicolumn{2}{c}{2270.748} && \multicolumn{2}{c}{5382.856} && \multicolumn{2}{c}{5367.720}&&  \multicolumn{2}{c}{5367.720} &&  \multicolumn{2}{c}{5367.720}&&\multicolumn{2}{c}{5367.720}\\
BIC &  \multicolumn{2}{c}{2305.720} && \multicolumn{2}{c}{5484.274} && \multicolumn{2}{c}{5493.618}&&  \multicolumn{2}{c}{5493.618} &&  \multicolumn{2}{c}{5493.618}&&\multicolumn{2}{c}{5493.618}\\
$R^2$ of Gf & \multicolumn{2}{c}{0.193}&& \multicolumn{2}{c}{N.A.} &&\multicolumn{2}{c}{0.193}&&  \multicolumn{2}{c}{0.193} &&  \multicolumn{2}{c}{0.193}&&\multicolumn{2}{c}{0.193}\\
\bottomrule
\end{longtable}
$^a$: In the one-step approach, the effects between the composites are no model parameters and were derived as indirect effects.
\end{landscape}

 Table~\ref{tab:resultsfree} reports the results of the three specifications with free weights.
 The blended and phantom-variable H–O specifications produce identical weight estimates. 
 For these approaches, the composite \textit{WMC} is formed as follows: $\textit{WMC}=0.116\cdot \textit{Ospa} + 1 \cdot \textit{Sspa}$.
 Thus, if \textit{Ospa} increases by one standard deviation without changing the other variables in the model, \textit{WMC} will increase by 0.166.
 While most of the SEs of the weights are the same, the SE of the weight of \textit{UR} is different. 
 This is likely because this weight is close to zero ($w_\text{UR}=0.021$).
 Consequently, the loading of $p_{31}$ on \textit{UR} which constitutes the basis for the weight (the loading equals the inverted weight), is rather large with a small standard error ($\lambda_\text{UR}=47.412$; $\text{SE}=3.407 \cdot 10^-5$). 
 As a result, the delta method, which is used by default in lavaan to derive estimates and SEs of transformed parameters, produces a very small SE for the weight of \textit{UR}. 
Moreover, the weights and SEs of the refined H–O specification differ from those of the other two approaches because a different scaling method was used.  
 In the refined H–O specification, one of the composite loadings is fixed to one, whereas in the blended and phantom-variable H–O specifications, one of the weights is fixed to one per composite of interest. 
 To facilitate comparison between the approaches, we report the standardized path coefficients and composite correlations, including their SEs. 
 As can be seen, the standardized parameters and SEs are identical across the approaches. 
 For example, if \textit{WMC} increases by one standard deviation without changing the other predictor composites, \textit{Gf} will increase by 0.236 standard deviations.
 Furthermore, the model fit statistics and information criteria are the same across the three approaches. 
 In contrast to the specifications in which the composites were modeled as averages, the three H–O specifications with free weights assume that all covariances between components of different composites are accounted for by the composites of interest.
 In other words, all effects of or on the composites' components are transmitted by the composites.
 For example, in the refined H–O specification, the effect of \textit{Ospa} on \textit{Gf} is $0.021 \approx 0.112 \cdot 0.184$, where 0.184 is the estimated unstandardized path coefficient of \textit{WMC} on \textit{Gf}.  
 This assumption imposes testable constraints on the variance-covariance matrix of the components.
 The various fit measures indicate a discrepancy between the model-implied and sample variance-covariance matrices of the components. 

When we compare the results with those in which the composites were modeled as averages, we see that more variance in \textit{Gf} can be explained when the weights are estimated freely. 
Similarly, the model information criteria indicate that the specifications with free weights are preferred over those with fixed weights.

\begin{table}[H]
\footnotesize
  \centering
\caption{Results for the illustrative example with freely estimated weights} \label{tab:resultsfree}
\begin{tabular}{lcccccccc}
\toprule
&\multicolumn{2}{p{2.9cm}}{\centering \pbox{20cm}{\relax\ifvmode\centering\fi Refined H–O\\specification}}&& 
\multicolumn{2}{p{2.9cm}}{\centering \pbox{20cm}{\relax\ifvmode\centering\fi Phantom-variable\\ H–O specification }}&& 
\multicolumn{2}{p{2.9cm}}{\centering \pbox{20cm}{\relax\ifvmode\centering\fi Blended H–O\\specification}}\\
\cline{2-3} \cline{5-6} \cline{8-9}  
& Est & SE && Est & SE && Est & SE\\[-12pt]
\textbf{Weights}&  &  &&  &  && & \\
$w_\text{Ospa}$ & 0.112 & 0.114 && 0.116 & 0.125 && 0.116 & 0.125\\
$w_\text{Sspa}$ & 0.962 & 0.052 && 1.000 & N.A.  && 1.000 & N.A.\\[1pt]
$w_\text{UE}$ & 0.303 & 0.100 && 0.372 & 0.168  && 0.372 & 0.168 \\
$w_\text{LE}$ & 0.816 & 0.099 && 1.000 & N.A.   && 1.000 & N.A. \\[1pt]
$w_\text{UR}$ & 0.021 & 0.173 && 0.021 & 0.000  && 0.021 & 0.176 \\
$w_\text{LR}$ & 0.992 & 0.071 && 1.000 & N.A.   && 1.000 & N.A. \\[1pt]
$w_\text{Lett}$ & 0.507 & 0.036  && 1.084 & 0.336 && 1.084 & 0.336\\
$w_\text{Rave}$ & 0.468 & 0.124  && 1.000 & N.A.  && 1.000 & N.A.\\[3pt]
\multicolumn{9}{l}{\textbf{Standardized path coefficients}}\\
$\gamma_1^\text{std}$: \textit{WMC} $\rightarrow$ \textit{Gf} & \phantom{-}0.236 & 0.062 && \phantom{-}0.236 & 0.062  &&  \phantom{-}0.236 & 0.062\\
$\gamma_2^\text{std}$: \textit{PE} $\rightarrow$ \textit{Gf} & -0.209 & 0.062 && -0.209 & 0.062  &&  -0.209 & 0.062\\
$\gamma_3^\text{std}$: \textit{PR} $\rightarrow$ \textit{Gf} & -0.292 & 0.056 && -0.292 & 0.056  &&  -0.292 & 0.056 \\[3pt]
\multicolumn{9}{l}{\textbf{Composite correlations}}\\
\textit{WMC} $\leftrightarrow$ \textit{PE} & -0.411 & 0.053 && -0.411 & 0.053  &&  -0.411 & 0.053 \\
\textit{WMC} $\leftrightarrow$ \textit{PR} & -0.120 & 0.063 && -0.120 & 0.063  &&  -0.120 & 0.063 \\
\textit{PE} $\leftrightarrow$ \textit{PR}  & -0.113 & 0.063 && -0.113 & 0.063   &&  -0.113 & 0.063 \\[3pt]
\multicolumn{9}{l}{\textbf{Fit statistics}}\\
$\chi^2$ statistic ($df$) &  \multicolumn{2}{c}{24.443 (14)}&& \multicolumn{2}{c}{24.443 (14)}  && \multicolumn{2}{c}{24.443 (14)}\\
RMSEA & \multicolumn{2}{c}{0.055} && \multicolumn{2}{c}{0.055} &&\multicolumn{2}{c}{0.055}\\
SRMR & \multicolumn{2}{c}{0.052} && \multicolumn{2}{c}{0.052}  &&\multicolumn{2}{c}{0.052}\\
AIC &  \multicolumn{2}{c}{5364.164} && \multicolumn{2}{c}{5364.164} &&\multicolumn{2}{c}{5364.164}\\
BIC &  \multicolumn{2}{c}{5441.101} && \multicolumn{2}{c}{5441.101} &&\multicolumn{2}{c}{5441.101}\\
$R^2$ of Gf & \multicolumn{2}{c}{0.228} &&\multicolumn{2}{c}{0.228}&&  \multicolumn{2}{c}{0.228}\\
\bottomrule
\end{tabular}
\end{table}

\section*{Discussion} \label{sec:discussion}

This paper provides an in-depth overview of existing approaches to studying composites in SEM, including their limitations. 
Among the existing approaches, the refined H–O specification is arguably the most versatile.
It can mimic the results of other approaches, such as the two-step, one-step, and pseudo-indicator approaches, as well as the original H–O specification, and it also overcomes their limitations. 
However, the refined H–O specification has limitations regarding the visibility the weights in the model specification and the ability to fix weights to specific values.
To address these issues, we propose two enhancements to the refined H–O specification: (i) the phantom-variable H–O specification and (ii) the blended H–O specification. 
In the phantom-variable H–O specification, the weights used to create the composite of interest are visible as the inverted loadings of the phantom variables on their respective components.
Consequently, researchers can easily identify the weights in graphical model representations or in the default output of  SEM software.
Additionally, this specification allows us to fix the weights very flexibly.
However, this specification may face limitations when the weights are close to zero; when a weight approaches zero, the respective loading of the phantom variable on the component approaches infinity.   
Consequently, it is not possible to fix weights to zero. 
Moreover, calculating SEs and drawing inferences about the estimated weights still requires an additional step: using the SEM software feature to specify the weights as the inverted loadings of the phantom variables on their respective components.
The delta method, which is based on asymptotics, is used by default for this purpose.  
In practice, it is often unclear whether the sample size is large enough for the asymptotic properties to hold. 
To address this issue, researchers can consider nonparametric approaches, such as bootstrap \citep{Efron1994} or jackknife \citep{Miller1974}, to draw inferences about the weights. 
The same has been suggested for making inferences about indirect effects \citep[see, e.g.,][]{Shrout2002}, such as the effects between the composites in our illustrative example using the modified one-step approach.
In contrast, in the blended H–O specification, the weights used to form the composite of interest are direct model parameters when the weight of the pseudo indicator is fixed to one.
Otherwise, the relevant parameters of the blended H–O specification represent the ratio of the actual weight to the weight of the pseudo indicator.  
Therefore, as with the phantom H–O specification, the weights can be easily fixed, even to zero.

Table \ref{tab:characteristics} juxtaposes the different approaches. 
The composite factor model specification and the MIMIC approach are excluded because the model is not identified and because it is unclear how to flexibly model composites, respectively.
The two-step approach is easy to apply, but ignores the composite creation and therefore misses opportunities for model assessment and missing data handling.
The one-step approach addresses this limitation by modeling the composite as a formatively measured latent variable with an error variance fixed to zero.
However, this approach has limitations when modeling composites as outcome variables in the structural model and when specifying covariances between composites (and other variables).
Consequently, it misses opportunities when testing whether a composite of interest fully transmits the effects of or on its components.
Similarly, this approach has limitations when relaxing this assumption.
Moreover, we consider a modification of the one-step approach.
While this modification enables us to model fixed-weight composites as outcome variables, the other limitations of the original one-step approach remain, as highlighted by our illustrative example.
Currently, it is generally not possible to replicate the results of the two-step approach. 
The pseudo-indicator approach allows us for flexible modeling of composites. 
However, it is only useful for modeling fixed-weight composites because it cannot model unknown-weight composites. 
Additionally, it is unclear how this approach can be used to test whether a composite fully transmits the effects on its components. 
The original and refined H–O specifications address almost all limitations of existing approaches. 
However, both approaches have limitations in their ability to study fixed-weight composites.
Our two newly proposed approaches -- the phantom-variable and blended H–O specifications -- maintain the advantages of the original and refined H–O specifications while addressing their limitations. 
Both approaches express weights more intuitively in the model specification, and allowing researchers to fix weights to specific values more flexibly.

\begin{landscape} 
\begin{table}
    \centering
        \caption{Characteristics of the different approaches to study composites in SEM}
    \label{tab:characteristics}
    \begin{tabular}{p{4cm}ccccccc}
    \toprule
    & \pbox{20cm}{Two-step\\ approach} &  \pbox{20cm}{One-step approach\\ including modification} &  \pbox{20cm}{Pseudo-indicator\\ approach} &    \pbox{20cm}{Original H–O \\specification}&  \pbox{20cm}{Refined H–O\\specification} &  \pbox{20cm}{Phantom-variable\\ H–O specification } & \pbox{20cm}{Blended H–O\\specification}\\
    \midrule
   Allows for...&&&&&&&\\ 
    ...taking into account the composite creation in the model. & \ding{55}& \ding{51}&\ding{51}& \ding{51} & \ding{51}  & \ding{51} & \ding{51}\\ \\ 
        ...identifying the weights in the model specification. &  N.A.  & \ding{51}  &   \ding{51}& \ding{55} & \ding{55} & \ding{51}  & \ding{51}\\ \\
    ...fixing composite weights to any value. &  \ding{51}  & \ding{51} & \ding{51} &   \ding{55} & \ding{109}  & \ding{109} & \ding{51}\\ \\
        ...freely estimating composite weights. &  \ding{55}  & \ding{51} & \ding{55} &   \ding{51} & \ding{51}  & \ding{51} & \ding{51}\\ \\
    ...modeling composites as outcome variables. &  \ding{51}  & \ding{109}& \ding{51}  &   \ding{51} & \ding{51}  & \ding{51} & \ding{51}\\ \\
    ...assessing whether the composites fully transmit the effects of or on their components.   &  \ding{55}  & \ding{109}& ?  &   \ding{51} & \ding{51}  & \ding{51} & \ding{51}\\ 
    \bottomrule
    \end{tabular}
    Note: \ding{51}: possible; \ding{109}: limitedly possible; \ding{55}: not possible; ?: currently not clear; N.A.: not applicable. 
    \end{table}
\end{landscape}

The introduction of the blended H–O specification does not only increase the flexibility of modeling composites in SEM, it can also help to answer more technical questions.
It strongly resembles a common factor model specification when the excrescent variables are considered as unique terms. 
Therefore, this specification can be helpful in investigating whether the common factor is (covariance) nested within the composite model \citep{Siegel2025}.
Similarly, this specification may be less prone to convergence issue during model estimation.
Since the blended H–O specification strongly resembles the common factor model specification, SEM software default starting values for common factor models may also be suitable for the blended H–O specification. 

In this paper, we focused only on approaches that are compatible with the LISREL SEM framework \citet{Joereskog1970a}, which is a special case of the reticular action model \citep{McArdle1984}. 
Consequently, these can be employed using common SEM software, such as Mplus \citep{MPLUS}, LISREL \citep{Joereskog2018} and the R packages  sem  \citep{Fox2024} and lavaan \citep{Rosseel2012}.
However, various approaches to estimating models containing composites have emerged outside this SEM framework, such as PLS-PM \citep[][]{Wold1982c} and GSCA \citep[][]{Hwang2004}.
Future research should compare these approaches further, including SEM composite specifications, such as the H–O specification and its enhancements, to better understand their similarities and differences \citep[see, e.g.,][for a recent comparison of PLS-PM, GSCA and regression component analysis]{Rigdon2024} . 
Moreover, in the context of PLS-PM and GSCA, researchers have begun distinguishing between two types of unknown-weight composites: nomological composites and canonical composites \citep{Cho2020,Cho2022c,Dijkstra2017a}.
The approaches to model unknown-weight estimates in our study rely on canonical composites, i.e., the weights are chosen in such a way that the sum of the explained variances of all dependent composites is maximized.
To bring SEM on par with GSCA and PLS-PM, future research must propose ways to model nomological composites in SEM, i.e., the weights are chosen in such a way that the sum of the explained variances of all components and dependent composites is maximized. 

\newpage
\bibliographystyle{apacite}
\bibliography{sample}

\begin{thebibliography}{}

\bibitem [\protect \citeauthoryear {%
Ahorsu%
\ \protect \BOthers {.}}{%
Ahorsu%
\ \protect \BOthers {.}}{%
{\protect \APACyear {2020}}%
}]{%
ahorsu2020}
\APACinsertmetastar {%
ahorsu2020}%
\begin{APACrefauthors}%
Ahorsu, D\BPBI K.%
, Imani, V.%
, Lin, C\BHBI Y.%
, Timpka, T.%
, Broström, A.%
, Updegraff, J\BPBI A.%
\BDBL {}Pakpour, A\BPBI H.%
\end{APACrefauthors}%
\unskip\
\newblock
\APACrefYearMonthDay{2020}{}{}.
\newblock
{\BBOQ}\APACrefatitle {Associations Between Fear of {COVID}-19, Mental Health,
  and Preventive Behaviours Across Pregnant Women and Husbands: An
  Actor-Partner Interdependence Modelling} {Associations between fear of
  {COVID}-19, mental health, and preventive behaviours across pregnant women
  and husbands: An actor-partner interdependence modelling}.{\BBCQ}
\newblock
\APACjournalVolNumPages{International Journal of Mental Health and
  Addiction}{20}{1}{68--82}.
\newblock
\begin{APACrefDOI} \doi{10.1007/s11469-020-00340-x} \end{APACrefDOI}
\PrintBackRefs{\CurrentBib}

\bibitem [\protect \citeauthoryear {%
Allison%
}{%
Allison%
}{%
{\protect \APACyear {2003}}%
}]{%
allison_2003}
\APACinsertmetastar {%
allison_2003}%
\begin{APACrefauthors}%
Allison, P\BPBI D.%
\end{APACrefauthors}%
\unskip\
\newblock
\APACrefYearMonthDay{2003}{}{}.
\newblock
{\BBOQ}\APACrefatitle {Missing Data Techniques for Structural Equation
  Modeling.} {Missing data techniques for structural equation modeling.}{\BBCQ}
\newblock
\APACjournalVolNumPages{Journal of Abnormal Psychology}{112}{4}{545–557}.
\newblock
\begin{APACrefDOI} \doi{https://doi.org/10.1037/0021-843x.112.4.545}
  \end{APACrefDOI}
\PrintBackRefs{\CurrentBib}

\bibitem [\protect \citeauthoryear {%
Bagozzi%
, Fornell%
\BCBL {}\ \BBA {} Larcker%
}{%
Bagozzi%
\ \protect \BOthers {.}}{%
{\protect \APACyear {1981}}%
}]{%
Bagozzi_Fornell_Larcker_1981}
\APACinsertmetastar {%
Bagozzi_Fornell_Larcker_1981}%
\begin{APACrefauthors}%
Bagozzi, R\BPBI P.%
, Fornell, C.%
\BCBL {}\ \BBA {} Larcker, D\BPBI F.%
\end{APACrefauthors}%
\unskip\
\newblock
\APACrefYearMonthDay{1981}{}{}.
\newblock
{\BBOQ}\APACrefatitle {Canonical correlation analysis as a special case of a
  structural relations model} {Canonical correlation analysis as a special case
  of a structural relations model}.{\BBCQ}
\newblock
\APACjournalVolNumPages{Multivariate Behavioral Research}{16}{4}{437--454}.
\newblock
\begin{APACrefDOI} \doi{10.1207/s15327906mbr1604_2} \end{APACrefDOI}
\PrintBackRefs{\CurrentBib}

\bibitem [\protect \citeauthoryear {%
Bandalos%
}{%
Bandalos%
}{%
{\protect \APACyear {2002}}%
}]{%
Bandalos2002}
\APACinsertmetastar {%
Bandalos2002}%
\begin{APACrefauthors}%
Bandalos, D\BPBI L.%
\end{APACrefauthors}%
\unskip\
\newblock
\APACrefYearMonthDay{2002}{}{}.
\newblock
{\BBOQ}\APACrefatitle {The Effects of Item Parceling on Goodness-of-Fit and
  Parameter Estimate Bias in Structural Equation Modeling} {The effects of item
  parceling on goodness-of-fit and parameter estimate bias in structural
  equation modeling}.{\BBCQ}
\newblock
\APACjournalVolNumPages{Structural Equation Modeling: A Multidisciplinary
  Journal}{9}{1}{78--102}.
\newblock
\begin{APACrefDOI} \doi{10.1207/s15328007sem0901_5} \end{APACrefDOI}
\PrintBackRefs{\CurrentBib}

\bibitem [\protect \citeauthoryear {%
Benitez%
, Llorens%
\BCBL {}\ \BBA {} Braojos%
}{%
Benitez%
\ \protect \BOthers {.}}{%
{\protect \APACyear {2018}}%
}]{%
Benitez2018b}
\APACinsertmetastar {%
Benitez2018b}%
\begin{APACrefauthors}%
Benitez, J.%
, Llorens, J.%
\BCBL {}\ \BBA {} Braojos, J.%
\end{APACrefauthors}%
\unskip\
\newblock
\APACrefYearMonthDay{2018}{}{}.
\newblock
{\BBOQ}\APACrefatitle {How information technology influences opportunity
  exploration and exploitation firm's capabilities} {How information technology
  influences opportunity exploration and exploitation firm's
  capabilities}.{\BBCQ}
\newblock
\APACjournalVolNumPages{Information {\&} Management}{55}{4}{508--523}.
\PrintBackRefs{\CurrentBib}

\bibitem [\protect \citeauthoryear {%
Bollen%
}{%
Bollen%
}{%
{\protect \APACyear {1989}}%
}]{%
bollen_1989}
\APACinsertmetastar {%
bollen_1989}%
\begin{APACrefauthors}%
Bollen, K\BPBI A.%
\end{APACrefauthors}%
\unskip\
\newblock
\APACrefYear{1989}.
\newblock
\APACrefbtitle {Structural equations with latent variables} {Structural
  equations with latent variables}.
\newblock
\APACaddressPublisher{New York}{Wiley}.
\PrintBackRefs{\CurrentBib}

\bibitem [\protect \citeauthoryear {%
Bollen%
\ \BBA {} Noble%
}{%
Bollen%
\ \BBA {} Noble%
}{%
{\protect \APACyear {2011}}%
}]{%
bollen_noble_2011}
\APACinsertmetastar {%
bollen_noble_2011}%
\begin{APACrefauthors}%
Bollen, K\BPBI A.%
\BCBT {}\ \BBA {} Noble, M\BPBI D.%
\end{APACrefauthors}%
\unskip\
\newblock
\APACrefYearMonthDay{2011}{}{}.
\newblock
{\BBOQ}\APACrefatitle {Structural equation models and the quantification of
  behavior} {Structural equation models and the quantification of
  behavior}.{\BBCQ}
\newblock
\APACjournalVolNumPages{Proceedings of the National Academy of
  Sciences}{108}{}{15639–15646}.
\newblock
\begin{APACrefDOI} \doi{https://doi.org/10.1073/pnas.1010661108}
  \end{APACrefDOI}
\PrintBackRefs{\CurrentBib}

\bibitem [\protect \citeauthoryear {%
Burgoyne%
, Hambrick%
\BCBL {}\ \BBA {} Altmann%
}{%
Burgoyne%
\ \protect \BOthers {.}}{%
{\protect \APACyear {2019}}%
}]{%
burgoyne_hambrick_altmann_2019}
\APACinsertmetastar {%
burgoyne_hambrick_altmann_2019}%
\begin{APACrefauthors}%
Burgoyne, A\BPBI P.%
, Hambrick, D\BPBI Z.%
\BCBL {}\ \BBA {} Altmann, E\BPBI M.%
\end{APACrefauthors}%
\unskip\
\newblock
\APACrefYearMonthDay{2019}{}{}.
\newblock
{\BBOQ}\APACrefatitle {Placekeeping Ability as a Component of Fluid
  Intelligence: Not Just Working Memory Capacity} {Placekeeping ability as a
  component of fluid intelligence: Not just working memory capacity}.{\BBCQ}
\newblock
\APACjournalVolNumPages{The American Journal of Psychology}{132}{4}{439–449}.
\newblock
\begin{APACrefDOI} \doi{https://doi.org/10.5406/amerjpsyc.132.4.0439}
  \end{APACrefDOI}
\PrintBackRefs{\CurrentBib}

\bibitem [\protect \citeauthoryear {%
Cadogan%
\ \BBA {} Lee%
}{%
Cadogan%
\ \BBA {} Lee%
}{%
{\protect \APACyear {2023}}%
{\protect \APACexlab {{\protect \BCnt {1}}}}}]{%
Cadogan2023}
\APACinsertmetastar {%
Cadogan2023}%
\begin{APACrefauthors}%
Cadogan, J\BPBI W.%
\BCBT {}\ \BBA {} Lee, N.%
\end{APACrefauthors}%
\unskip\
\newblock
\APACrefYearMonthDay{2023{\protect \BCnt {1}}}{}{}.
\newblock
{\BBOQ}\APACrefatitle {A Miracle of Measurement or Accidental Constructivism?
  {H}ow {PLS} Subverts the Realist Search for Truth} {A miracle of measurement
  or accidental constructivism? {H}ow {PLS} subverts the realist search for
  truth}.{\BBCQ}
\newblock
\APACjournalVolNumPages{European Journal of Marketing}{57}{6}{1703--1724}.
\newblock
\begin{APACrefDOI} \doi{10.1108/EJM-08-2020-0637} \end{APACrefDOI}
\PrintBackRefs{\CurrentBib}

\bibitem [\protect \citeauthoryear {%
Cadogan%
\ \BBA {} Lee%
}{%
Cadogan%
\ \BBA {} Lee%
}{%
{\protect \APACyear {2023}}%
{\protect \APACexlab {{\protect \BCnt {2}}}}}]{%
Cadogan2023a}
\APACinsertmetastar {%
Cadogan2023a}%
\begin{APACrefauthors}%
Cadogan, J\BPBI W.%
\BCBT {}\ \BBA {} Lee, N.%
\end{APACrefauthors}%
\unskip\
\newblock
\APACrefYearMonthDay{2023{\protect \BCnt {2}}}{}{}.
\newblock
{\BBOQ}\APACrefatitle {Scientific realism, the necessity of causal contact in
  measurement and emergent variables} {Scientific realism, the necessity of
  causal contact in measurement and emergent variables}.{\BBCQ}
\newblock
\APACjournalVolNumPages{European Journal of Marketing}{57}{6}{1758--1779}.
\newblock
\begin{APACrefDOI} \doi{10.1108/ejm-06-2021-0454} \end{APACrefDOI}
\PrintBackRefs{\CurrentBib}

\bibitem [\protect \citeauthoryear {%
Cho%
\ \BBA {} Choi%
}{%
Cho%
\ \BBA {} Choi%
}{%
{\protect \APACyear {2020}}%
}]{%
Cho2020}
\APACinsertmetastar {%
Cho2020}%
\begin{APACrefauthors}%
Cho, G.%
\BCBT {}\ \BBA {} Choi, J\BPBI Y.%
\end{APACrefauthors}%
\unskip\
\newblock
\APACrefYearMonthDay{2020}{}{}.
\newblock
{\BBOQ}\APACrefatitle {An empirical comparison of generalized structured
  component analysis and partial least squares path modeling under
  variance-based structural equation models} {An empirical comparison of
  generalized structured component analysis and partial least squares path
  modeling under variance-based structural equation models}.{\BBCQ}
\newblock
\APACjournalVolNumPages{Behaviormetrika}{47}{}{243--272}.
\newblock
\begin{APACrefDOI} \doi{10.1007/s41237-019-00098-0} \end{APACrefDOI}
\PrintBackRefs{\CurrentBib}

\bibitem [\protect \citeauthoryear {%
Cho%
, Hwang%
\BCBL {}\ \BBA {} Sarstedt%
}{%
Cho%
\ \protect \BOthers {.}}{%
{\protect \APACyear {2022}}%
}]{%
Cho2022c}
\APACinsertmetastar {%
Cho2022c}%
\begin{APACrefauthors}%
Cho, G.%
, Hwang, H.%
\BCBL {}\ \BBA {} Sarstedt, M.%
\end{APACrefauthors}%
\unskip\
\newblock
\APACrefYearMonthDay{2022}{}{}.
\newblock
{\BBOQ}\APACrefatitle {A comparative evaluation of factor- and component-based
  structural equation modeling approaches under (in)correct construct
  representations} {A comparative evaluation of factor- and component-based
  structural equation modeling approaches under (in)correct construct
  representations}.{\BBCQ}
\newblock
\APACjournalVolNumPages{British Journal of Mathematical and Statistical
  Psychology}{75}{2}{220-251}.
\newblock
\begin{APACrefDOI} \doi{10.1111/bmsp.12255} \end{APACrefDOI}
\PrintBackRefs{\CurrentBib}

\bibitem [\protect \citeauthoryear {%
Dijkstra%
}{%
Dijkstra%
}{%
{\protect \APACyear {2017}}%
}]{%
Dijkstra2017a}
\APACinsertmetastar {%
Dijkstra2017a}%
\begin{APACrefauthors}%
Dijkstra, T\BPBI K.%
\end{APACrefauthors}%
\unskip\
\newblock
\APACrefYearMonthDay{2017}{}{}.
\newblock
{\BBOQ}\APACrefatitle {A perfect match between a model and a mode} {A perfect
  match between a model and a mode}.{\BBCQ}
\newblock
\BIn{} H.~Latan\ \BBA {} R.~Noonan\ (\BEDS), \APACrefbtitle {Partial Least
  Squares Path Modeling: Basic Concepts, Methodological Issues and
  Applications} {Partial least squares path modeling: Basic concepts,
  methodological issues and applications}\ (\BPG~55-80).
\newblock
\APACaddressPublisher{Cham}{Springer}.
\PrintBackRefs{\CurrentBib}

\bibitem [\protect \citeauthoryear {%
Edwards%
}{%
Edwards%
}{%
{\protect \APACyear {2001}}%
}]{%
Edwards2001}
\APACinsertmetastar {%
Edwards2001}%
\begin{APACrefauthors}%
Edwards, J\BPBI R.%
\end{APACrefauthors}%
\unskip\
\newblock
\APACrefYearMonthDay{2001}{}{}.
\newblock
{\BBOQ}\APACrefatitle {Multidimensional constructs in organizational behavior
  research: An integrative analytical framework} {Multidimensional constructs
  in organizational behavior research: An integrative analytical
  framework}.{\BBCQ}
\newblock
\APACjournalVolNumPages{Organizational Research Methods}{4}{2}{144--192}.
\newblock
\begin{APACrefDOI} \doi{10.1177/109442810142004} \end{APACrefDOI}
\PrintBackRefs{\CurrentBib}

\bibitem [\protect \citeauthoryear {%
Efron%
\ \BBA {} Tibshirani%
}{%
Efron%
\ \BBA {} Tibshirani%
}{%
{\protect \APACyear {1994}}%
}]{%
Efron1994}
\APACinsertmetastar {%
Efron1994}%
\begin{APACrefauthors}%
Efron, B.%
\BCBT {}\ \BBA {} Tibshirani, R\BPBI J.%
\end{APACrefauthors}%
\unskip\
\newblock
\APACrefYear{1994}.
\newblock
\APACrefbtitle {An Introduction to the Bootstrap} {An introduction to the
  bootstrap}.
\newblock
\APACaddressPublisher{Boca Raton}{Chapman \& Hall/CRC}.
\PrintBackRefs{\CurrentBib}

\bibitem [\protect \citeauthoryear {%
Fox%
, Nie%
\BCBL {}\ \BBA {} Byrnes%
}{%
Fox%
\ \protect \BOthers {.}}{%
{\protect \APACyear {2024}}%
}]{%
Fox2024}
\APACinsertmetastar {%
Fox2024}%
\begin{APACrefauthors}%
Fox, J.%
, Nie, Z.%
\BCBL {}\ \BBA {} Byrnes, J.%
\end{APACrefauthors}%
\unskip\
\newblock
\APACrefYearMonthDay{2024}{}{}.
\newblock
{\BBOQ}\APACrefatitle {sem: Structural Equation Models} {sem: Structural
  equation models}{\BBCQ}\ [\bibcomputersoftwaremanual].
\newblock
\begin{APACrefURL} \url{https://CRAN.R-project.org/package=sem}
  \end{APACrefURL}
\newblock
\APACrefnote{R package version 3.1-16}
\newblock
\begin{APACrefDOI} \doi{10.32614/CRAN.package.sem} \end{APACrefDOI}
\PrintBackRefs{\CurrentBib}

\bibitem [\protect \citeauthoryear {%
Grace%
\ \BBA {} Bollen%
}{%
Grace%
\ \BBA {} Bollen%
}{%
{\protect \APACyear {2008}}%
}]{%
grace_bollen_2008}
\APACinsertmetastar {%
grace_bollen_2008}%
\begin{APACrefauthors}%
Grace, J\BPBI B.%
\BCBT {}\ \BBA {} Bollen, K\BPBI A.%
\end{APACrefauthors}%
\unskip\
\newblock
\APACrefYearMonthDay{2008}{}{}.
\newblock
{\BBOQ}\APACrefatitle {Representing general theoretical concepts in structural
  equation models: The role of composite variables} {Representing general
  theoretical concepts in structural equation models: The role of composite
  variables}.{\BBCQ}
\newblock
\APACjournalVolNumPages{Environmental and Ecological
  Statistics}{15}{2}{191--213}.
\newblock
\begin{APACrefDOI} \doi{10.1007/s10651-007-0047-7} \end{APACrefDOI}
\PrintBackRefs{\CurrentBib}

\bibitem [\protect \citeauthoryear {%
Graham%
}{%
Graham%
}{%
{\protect \APACyear {2008}}%
}]{%
Graham2008}
\APACinsertmetastar {%
Graham2008}%
\begin{APACrefauthors}%
Graham, J\BPBI M.%
\end{APACrefauthors}%
\unskip\
\newblock
\APACrefYearMonthDay{2008}{}{}.
\newblock
{\BBOQ}\APACrefatitle {The General Linear Model as Structural Equation
  Modeling} {The general linear model as structural equation modeling}.{\BBCQ}
\newblock
\APACjournalVolNumPages{Journal of Educational and Behavioral
  Statistics}{33}{4}{485--506}.
\newblock
\begin{APACrefDOI} \doi{10.3102/1076998607306151} \end{APACrefDOI}
\PrintBackRefs{\CurrentBib}

\bibitem [\protect \citeauthoryear {%
Heise%
}{%
Heise%
}{%
{\protect \APACyear {1972}}%
}]{%
Heise1972}
\APACinsertmetastar {%
Heise1972}%
\begin{APACrefauthors}%
Heise, D\BPBI R.%
\end{APACrefauthors}%
\unskip\
\newblock
\APACrefYearMonthDay{1972}{}{}.
\newblock
{\BBOQ}\APACrefatitle {Employing Nominal Variables, Induced Variables, and
  Block Variables in Path Analyses} {Employing nominal variables, induced
  variables, and block variables in path analyses}.{\BBCQ}
\newblock
\APACjournalVolNumPages{Sociological Methods {\&} Research}{1}{2}{147--173}.
\newblock
\begin{APACrefDOI} \doi{10.1177/004912417200100201} \end{APACrefDOI}
\PrintBackRefs{\CurrentBib}

\bibitem [\protect \citeauthoryear {%
Henseler%
}{%
Henseler%
}{%
{\protect \APACyear {2021}}%
}]{%
henseler_2021}
\APACinsertmetastar {%
henseler_2021}%
\begin{APACrefauthors}%
Henseler, J.%
\end{APACrefauthors}%
\unskip\
\newblock
\APACrefYear{2021}.
\newblock
\APACrefbtitle {Composite-based structural equation modeling: {A}nalyzing
  latent and emergent variables} {Composite-based structural equation modeling:
  {A}nalyzing latent and emergent variables}.
\newblock
\APACaddressPublisher{New York}{Guilford Press}.
\PrintBackRefs{\CurrentBib}

\bibitem [\protect \citeauthoryear {%
Henseler%
\ \protect \BOthers {.}}{%
Henseler%
\ \protect \BOthers {.}}{%
{\protect \APACyear {2014}}%
}]{%
Henseler2014}
\APACinsertmetastar {%
Henseler2014}%
\begin{APACrefauthors}%
Henseler, J.%
, Dijkstra, T\BPBI K.%
, Sarstedt, M.%
, Ringle, C\BPBI M.%
, Diamantopoulos, A.%
, Straub, D\BPBI W.%
\BDBL {}Calantone, R\BPBI J.%
\end{APACrefauthors}%
\unskip\
\newblock
\APACrefYearMonthDay{2014}{}{}.
\newblock
{\BBOQ}\APACrefatitle {Common beliefs and reality about {PLS}: {C}omments on
  {R{\"o}nkk{\"o}} and {Evermann} (2013)} {Common beliefs and reality about
  {PLS}: {C}omments on {R{\"o}nkk{\"o}} and {Evermann} (2013)}.{\BBCQ}
\newblock
\APACjournalVolNumPages{Organizational Research Methods}{17}{2}{182--209}.
\newblock
\begin{APACrefDOI} \doi{10.1177/1094428114526928} \end{APACrefDOI}
\PrintBackRefs{\CurrentBib}

\bibitem [\protect \citeauthoryear {%
Henseler%
\ \BBA {} Schuberth%
}{%
Henseler%
\ \BBA {} Schuberth%
}{%
{\protect \APACyear {2021}}%
}]{%
Henseler2021}
\APACinsertmetastar {%
Henseler2021}%
\begin{APACrefauthors}%
Henseler, J.%
\BCBT {}\ \BBA {} Schuberth, F.%
\end{APACrefauthors}%
\unskip\
\newblock
\APACrefYearMonthDay{2021}{}{}.
\newblock
{\BBOQ}\APACrefatitle {Auxiliary Theories} {Auxiliary theories}.{\BBCQ}
\newblock
\BIn{} J.~Henseler\ (\BED), \APACrefbtitle {Composite-based Structural Equation
  Modeling: Analyzing Latent and Emergent Variables} {Composite-based
  structural equation modeling: Analyzing latent and emergent variables}\
  (\BPGS\ 25--37).
\newblock
\APACaddressPublisher{London, New York}{The Guilford Press}.
\PrintBackRefs{\CurrentBib}

\bibitem [\protect \citeauthoryear {%
Hwang%
\ \BBA {} Takane%
}{%
Hwang%
\ \BBA {} Takane%
}{%
{\protect \APACyear {2004}}%
}]{%
Hwang2004}
\APACinsertmetastar {%
Hwang2004}%
\begin{APACrefauthors}%
Hwang, H.%
\BCBT {}\ \BBA {} Takane, Y.%
\end{APACrefauthors}%
\unskip\
\newblock
\APACrefYearMonthDay{2004}{}{}.
\newblock
{\BBOQ}\APACrefatitle {Generalized structured component analysis} {Generalized
  structured component analysis}.{\BBCQ}
\newblock
\APACjournalVolNumPages{Psychometrika}{69}{1}{81--99}.
\PrintBackRefs{\CurrentBib}

\bibitem [\protect \citeauthoryear {%
J\"oreskog%
\ \BBA {} S\"orbom%
}{%
J\"oreskog%
\ \BBA {} S\"orbom%
}{%
{\protect \APACyear {2018}}%
}]{%
Joereskog2018}
\APACinsertmetastar {%
Joereskog2018}%
\begin{APACrefauthors}%
J\"oreskog, K.%
\BCBT {}\ \BBA {} S\"orbom, D.%
\end{APACrefauthors}%
\unskip\
\newblock
\APACrefYear{2018}.
\newblock
\APACrefbtitle {{LISREL} 10 for {W}indows} {{LISREL} 10 for {W}indows}.
\newblock
\APACaddressPublisher{Skokie, IL}{Scientific Software International, Inc}.
\PrintBackRefs{\CurrentBib}

\bibitem [\protect \citeauthoryear {%
J{\"o}reskog%
}{%
J{\"o}reskog%
}{%
{\protect \APACyear {1970}}%
}]{%
Joereskog1970a}
\APACinsertmetastar {%
Joereskog1970a}%
\begin{APACrefauthors}%
J{\"o}reskog, K\BPBI G.%
\end{APACrefauthors}%
\unskip\
\newblock
\APACrefYearMonthDay{1970}{}{}.
\newblock
{\BBOQ}\APACrefatitle {A general method for analysis of covariance structures}
  {A general method for analysis of covariance structures}.{\BBCQ}
\newblock
\APACjournalVolNumPages{Biometrika}{57}{2}{239--251}.
\newblock
\begin{APACrefDOI} \doi{10.2307/2334833} \end{APACrefDOI}
\PrintBackRefs{\CurrentBib}

\bibitem [\protect \citeauthoryear {%
J\"oreskog%
\ \BBA {} Goldberger%
}{%
J\"oreskog%
\ \BBA {} Goldberger%
}{%
{\protect \APACyear {1975}}%
}]{%
Joreskog1975}
\APACinsertmetastar {%
Joreskog1975}%
\begin{APACrefauthors}%
J\"oreskog, K\BPBI G.%
\BCBT {}\ \BBA {} Goldberger, A\BPBI S.%
\end{APACrefauthors}%
\unskip\
\newblock
\APACrefYearMonthDay{1975}{}{}.
\newblock
{\BBOQ}\APACrefatitle {Estimation of a Model with Multiple Indicators and
  Multiple Causes of a Single Latent Variable} {Estimation of a model with
  multiple indicators and multiple causes of a single latent variable}.{\BBCQ}
\newblock
\APACjournalVolNumPages{Journal of the American Statistical
  Association}{70}{351}{631--639}.
\newblock
\begin{APACrefDOI} \doi{10.2307/2285946} \end{APACrefDOI}
\PrintBackRefs{\CurrentBib}

\bibitem [\protect \citeauthoryear {%
Little%
}{%
Little%
}{%
{\protect \APACyear {2024}}%
}]{%
little2024}
\APACinsertmetastar {%
little2024}%
\begin{APACrefauthors}%
Little, T\BPBI D.%
\end{APACrefauthors}%
\unskip\
\newblock
\APACrefYear{2024}.
\newblock
\APACrefbtitle {Longitudinal Structural Equation Modeling} {Longitudinal
  structural equation modeling}.
\newblock
\APACaddressPublisher{New York}{Guilford Press}.
\PrintBackRefs{\CurrentBib}

\bibitem [\protect \citeauthoryear {%
Little%
, Rhemtulla%
, Gibson%
\BCBL {}\ \BBA {} Schoemann%
}{%
Little%
\ \protect \BOthers {.}}{%
{\protect \APACyear {2013}}%
}]{%
Little2013a}
\APACinsertmetastar {%
Little2013a}%
\begin{APACrefauthors}%
Little, T\BPBI D.%
, Rhemtulla, M.%
, Gibson, K.%
\BCBL {}\ \BBA {} Schoemann, A\BPBI M.%
\end{APACrefauthors}%
\unskip\
\newblock
\APACrefYearMonthDay{2013}{}{}.
\newblock
{\BBOQ}\APACrefatitle {Why the items versus parcels controversy needn't be one}
  {Why the items versus parcels controversy needn't be one}.{\BBCQ}
\newblock
\APACjournalVolNumPages{Psychological Methods}{18}{3}{285--300}.
\PrintBackRefs{\CurrentBib}

\bibitem [\protect \citeauthoryear {%
MacCallum%
\ \BBA {} Browne%
}{%
MacCallum%
\ \BBA {} Browne%
}{%
{\protect \APACyear {1993}}%
}]{%
MacCallum1993}
\APACinsertmetastar {%
MacCallum1993}%
\begin{APACrefauthors}%
MacCallum, R\BPBI C.%
\BCBT {}\ \BBA {} Browne, M\BPBI W.%
\end{APACrefauthors}%
\unskip\
\newblock
\APACrefYearMonthDay{1993}{}{}.
\newblock
{\BBOQ}\APACrefatitle {The use of causal indicators in covariance structure
  models: {S}ome practical issues.} {The use of causal indicators in covariance
  structure models: {S}ome practical issues.}{\BBCQ}
\newblock
\APACjournalVolNumPages{Psychological Bulletin}{114}{3}{533--541}.
\PrintBackRefs{\CurrentBib}

\bibitem [\protect \citeauthoryear {%
McArdle%
\ \BBA {} McDonald%
}{%
McArdle%
\ \BBA {} McDonald%
}{%
{\protect \APACyear {1984}}%
}]{%
McArdle1984}
\APACinsertmetastar {%
McArdle1984}%
\begin{APACrefauthors}%
McArdle, J\BPBI J.%
\BCBT {}\ \BBA {} McDonald, R\BPBI P.%
\end{APACrefauthors}%
\unskip\
\newblock
\APACrefYearMonthDay{1984}{}{}.
\newblock
{\BBOQ}\APACrefatitle {Some algebraic properties of the Reticular Action Model
  for moment structures} {Some algebraic properties of the reticular action
  model for moment structures}.{\BBCQ}
\newblock
\APACjournalVolNumPages{British Journal of Mathematical and Statistical
  Psychology}{37}{2}{234--251}.
\newblock
\begin{APACrefDOI} \doi{10.1111/j.2044-8317.1984.tb00802.x} \end{APACrefDOI}
\PrintBackRefs{\CurrentBib}

\bibitem [\protect \citeauthoryear {%
McCormick%
, Borgeest%
\BCBL {}\ \BBA {} Kievit%
}{%
McCormick%
\ \protect \BOthers {.}}{%
{\protect \APACyear {2022}}%
}]{%
mccormick_borgeest_kievit_2022}
\APACinsertmetastar {%
mccormick_borgeest_kievit_2022}%
\begin{APACrefauthors}%
McCormick, E.%
, Borgeest, G.%
\BCBL {}\ \BBA {} Kievit, R.%
\end{APACrefauthors}%
\unskip\
\newblock
\APACrefYearMonthDay{2022}{}{}.
\newblock
{\BBOQ}\APACrefatitle {Interrupted mediation: A cautionary note on using
  derived metrics as intervening variables in path models} {Interrupted
  mediation: A cautionary note on using derived metrics as intervening
  variables in path models}.{\BBCQ}
\newblock
\APACjournalVolNumPages{OSF}{}{}{}.
\newblock
\begin{APACrefURL} \url{https://osf.io/preprints/psyarxiv/48xj5}
  \end{APACrefURL}
\PrintBackRefs{\CurrentBib}

\bibitem [\protect \citeauthoryear {%
McIntosh%
, Edwards%
\BCBL {}\ \BBA {} Antonakis%
}{%
McIntosh%
\ \protect \BOthers {.}}{%
{\protect \APACyear {2014}}%
}]{%
mcintosh_edwards_antonakis_2014}
\APACinsertmetastar {%
mcintosh_edwards_antonakis_2014}%
\begin{APACrefauthors}%
McIntosh, C\BPBI N.%
, Edwards, J\BPBI R.%
\BCBL {}\ \BBA {} Antonakis, J.%
\end{APACrefauthors}%
\unskip\
\newblock
\APACrefYearMonthDay{2014}{}{}.
\newblock
{\BBOQ}\APACrefatitle {Reflections on Partial Least Squares Path Modeling}
  {Reflections on partial least squares path modeling}.{\BBCQ}
\newblock
\APACjournalVolNumPages{Organizational Research Methods}{17}{2}{210–251}.
\newblock
\begin{APACrefDOI} \doi{10.1177/1094428114529165} \end{APACrefDOI}
\PrintBackRefs{\CurrentBib}

\bibitem [\protect \citeauthoryear {%
Miller%
}{%
Miller%
}{%
{\protect \APACyear {1974}}%
}]{%
Miller1974}
\APACinsertmetastar {%
Miller1974}%
\begin{APACrefauthors}%
Miller, R\BPBI G.%
\end{APACrefauthors}%
\unskip\
\newblock
\APACrefYearMonthDay{1974}{}{}.
\newblock
{\BBOQ}\APACrefatitle {The jackknife -- a review} {The jackknife -- a
  review}.{\BBCQ}
\newblock
\APACjournalVolNumPages{Biometrika}{61}{1}{1--15}.
\PrintBackRefs{\CurrentBib}

\bibitem [\protect \citeauthoryear {%
Muth{\'e}n%
\ \BBA {} Muth{\'e}n%
}{%
Muth{\'e}n%
\ \BBA {} Muth{\'e}n%
}{%
{\protect \APACyear {1998-2017}}%
}]{%
MPLUS}
\APACinsertmetastar {%
MPLUS}%
\begin{APACrefauthors}%
Muth{\'e}n, L\BPBI K.%
\BCBT {}\ \BBA {} Muth{\'e}n, B\BPBI O.%
\end{APACrefauthors}%
\unskip\
\newblock
\APACrefYear{1998-2017}.
\newblock
\APACrefbtitle {Mplus} {Mplus}\ (\PrintOrdinal{8th}\ \BEd).
\newblock
\APACaddressPublisher{Los Angeles, CA}{Muth{\'e}n \& Muth{\'e}n}.
\PrintBackRefs{\CurrentBib}

\bibitem [\protect \citeauthoryear {%
Ogasawara%
}{%
Ogasawara%
}{%
{\protect \APACyear {2007}}%
}]{%
Ogasawara2007}
\APACinsertmetastar {%
Ogasawara2007}%
\begin{APACrefauthors}%
Ogasawara, H.%
\end{APACrefauthors}%
\unskip\
\newblock
\APACrefYearMonthDay{2007}{}{}.
\newblock
{\BBOQ}\APACrefatitle {Asymptotic expansions of the distributions of estimators
  in canonical correlation analysis under nonnormality} {Asymptotic expansions
  of the distributions of estimators in canonical correlation analysis under
  nonnormality}.{\BBCQ}
\newblock
\APACjournalVolNumPages{Journal of Multivariate Analysis}{98}{9}{1726--1750}.
\newblock
\begin{APACrefDOI} \doi{10.1016/j.jmva.2006.12.001} \end{APACrefDOI}
\PrintBackRefs{\CurrentBib}

\bibitem [\protect \citeauthoryear {%
Pan%
\ \protect \BOthers {.}}{%
Pan%
\ \protect \BOthers {.}}{%
{\protect \APACyear {2023}}%
}]{%
pan_zhang_wang_zhao_zhao_ding_li_miao_fan_2023}
\APACinsertmetastar {%
pan_zhang_wang_zhao_zhao_ding_li_miao_fan_2023}%
\begin{APACrefauthors}%
Pan, L.%
, Zhang, X.%
, Wang, S.%
, Zhao, N.%
, Zhao, R.%
, Ding, B.%
\BDBL {}Fan, H.%
\end{APACrefauthors}%
\unskip\
\newblock
\APACrefYearMonthDay{2023}{}{}.
\newblock
{\BBOQ}\APACrefatitle {Determinants associated with self-management behavior
  among type 2 diabetes patients in {C}hina: A structural equation model based
  on the theory of planned behavior} {Determinants associated with
  self-management behavior among type 2 diabetes patients in {C}hina: A
  structural equation model based on the theory of planned behavior}.{\BBCQ}
\newblock
\APACjournalVolNumPages{International Journal of Clinical and Health
  Psychology}{23}{1}{100332}.
\newblock
\begin{APACrefDOI} \doi{https://doi.org/10.1016/j.ijchp.2022.100332}
  \end{APACrefDOI}
\PrintBackRefs{\CurrentBib}

\bibitem [\protect \citeauthoryear {%
{R Core Team}%
}{%
{R Core Team}%
}{%
{\protect \APACyear {2024}}%
}]{%
Rsoftware}
\APACinsertmetastar {%
Rsoftware}%
\begin{APACrefauthors}%
{R Core Team}.%
\end{APACrefauthors}%
\unskip\
\newblock
\APACrefYearMonthDay{2024}{}{}.
\newblock
{\BBOQ}\APACrefatitle {{R}: A Language and Environment for Statistical
  Computing} {{R}: A language and environment for statistical
  computing}{\BBCQ}\ [\bibcomputersoftwaremanual].
\newblock
\APACaddressPublisher{Vienna, Austria}{}.
\newblock
\begin{APACrefURL} \url{https://www.R-project.org/} \end{APACrefURL}
\PrintBackRefs{\CurrentBib}

\bibitem [\protect \citeauthoryear {%
Rasoolimanesh%
, Ali%
, Mikuli{\'{c}}%
\BCBL {}\ \BBA {} Dogan%
}{%
Rasoolimanesh%
\ \protect \BOthers {.}}{%
{\protect \APACyear {2023}}%
}]{%
Rasoolimanesh2023}
\APACinsertmetastar {%
Rasoolimanesh2023}%
\begin{APACrefauthors}%
Rasoolimanesh, S\BPBI M.%
, Ali, F.%
, Mikuli{\'{c}}, J.%
\BCBL {}\ \BBA {} Dogan, S.%
\end{APACrefauthors}%
\unskip\
\newblock
\APACrefYearMonthDay{2023}{}{}.
\newblock
{\BBOQ}\APACrefatitle {Reflective and composite scales in tourism and
  hospitality research: Revising the scale development procedure} {Reflective
  and composite scales in tourism and hospitality research: Revising the scale
  development procedure}.{\BBCQ}
\newblock
\APACjournalVolNumPages{International Journal of Contemporary Hospitality
  Management}{35}{2}{590--601}.
\newblock
\begin{APACrefDOI} \doi{10.1108/ijchm-02-2022-0255} \end{APACrefDOI}
\PrintBackRefs{\CurrentBib}

\bibitem [\protect \citeauthoryear {%
Rigdon%
}{%
Rigdon%
}{%
{\protect \APACyear {2012}}%
}]{%
Rigdon2012}
\APACinsertmetastar {%
Rigdon2012}%
\begin{APACrefauthors}%
Rigdon, E\BPBI E.%
\end{APACrefauthors}%
\unskip\
\newblock
\APACrefYearMonthDay{2012}{}{}.
\newblock
{\BBOQ}\APACrefatitle {Rethinking partial least squares path modeling: {I}n
  praise of simple methods} {Rethinking partial least squares path modeling:
  {I}n praise of simple methods}.{\BBCQ}
\newblock
\APACjournalVolNumPages{Long Range Planning}{45}{5}{341--358}.
\newblock
\begin{APACrefDOI} \doi{10.1016/j.lrp.2012.09.010} \end{APACrefDOI}
\PrintBackRefs{\CurrentBib}

\bibitem [\protect \citeauthoryear {%
Rigdon%
}{%
Rigdon%
}{%
{\protect \APACyear {2023}}%
}]{%
Rigdon2023d}
\APACinsertmetastar {%
Rigdon2023d}%
\begin{APACrefauthors}%
Rigdon, E\BPBI E.%
\end{APACrefauthors}%
\unskip\
\newblock
\APACrefYearMonthDay{2023}{}{}.
\newblock
{\BBOQ}\APACrefatitle {The proxy of {D}orian {G}ray: Scientific realism,
  construct validation and the way forward} {The proxy of {D}orian {G}ray:
  Scientific realism, construct validation and the way forward}.{\BBCQ}
\newblock
\APACjournalVolNumPages{European Journal of Marketing}{57}{6}{1725--1736}.
\newblock
\begin{APACrefDOI} \doi{10.1108/ejm-05-2021-0319} \end{APACrefDOI}
\PrintBackRefs{\CurrentBib}

\bibitem [\protect \citeauthoryear {%
Rigdon%
}{%
Rigdon%
}{%
{\protect \APACyear {2024}}%
}]{%
Rigdon2024}
\APACinsertmetastar {%
Rigdon2024}%
\begin{APACrefauthors}%
Rigdon, E\BPBI E.%
\end{APACrefauthors}%
\unskip\
\newblock
\APACrefYearMonthDay{2024}{}{}.
\newblock
{\BBOQ}\APACrefatitle {Understanding Composite-Based Structural Equation
  Modeling Methods Fromthe Perspective of Regression Component Analysis}
  {Understanding composite-based structural equation modeling methods fromthe
  perspective of regression component analysis}.{\BBCQ}
\newblock
\APACjournalVolNumPages{Multivariate Behavioral Research}{59}{4}{677--692}.
\PrintBackRefs{\CurrentBib}

\bibitem [\protect \citeauthoryear {%
Rose%
, Wagner%
, Mayer%
\BCBL {}\ \BBA {} Nagengast%
}{%
Rose%
\ \protect \BOthers {.}}{%
{\protect \APACyear {2019}}%
}]{%
Rose_Wagner_Mayer_Nagengast_2019}
\APACinsertmetastar {%
Rose_Wagner_Mayer_Nagengast_2019}%
\begin{APACrefauthors}%
Rose, N.%
, Wagner, W.%
, Mayer, A.%
\BCBL {}\ \BBA {} Nagengast, B.%
\end{APACrefauthors}%
\unskip\
\newblock
\APACrefYearMonthDay{2019}{}{}.
\newblock
{\BBOQ}\APACrefatitle {Model-based manifest and latent composite scores in
  structural equation models} {Model-based manifest and latent composite scores
  in structural equation models}.{\BBCQ}
\newblock
\APACjournalVolNumPages{Collabra: Psychology}{5}{1}{9}.
\newblock
\begin{APACrefDOI} \doi{10.1525/collabra.143} \end{APACrefDOI}
\PrintBackRefs{\CurrentBib}

\bibitem [\protect \citeauthoryear {%
Rosseel%
}{%
Rosseel%
}{%
{\protect \APACyear {2012}}%
}]{%
Rosseel2012}
\APACinsertmetastar {%
Rosseel2012}%
\begin{APACrefauthors}%
Rosseel, Y.%
\end{APACrefauthors}%
\unskip\
\newblock
\APACrefYearMonthDay{2012}{}{}.
\newblock
{\BBOQ}\APACrefatitle {lavaan: {A}n {R} package for structural equation
  modeling} {lavaan: {A}n {R} package for structural equation modeling}.{\BBCQ}
\newblock
\APACjournalVolNumPages{Journal of Statistical Software}{48}{2}{1--36}.
\PrintBackRefs{\CurrentBib}

\bibitem [\protect \citeauthoryear {%
Rosseel%
, Jorgensen%
\BCBL {}\ \BBA {} {De Wilde}%
}{%
Rosseel%
\ \protect \BOthers {.}}{%
{\protect \APACyear {2025}}%
}]{%
lavaan}
\APACinsertmetastar {%
lavaan}%
\begin{APACrefauthors}%
Rosseel, Y.%
, Jorgensen, T\BPBI D.%
\BCBL {}\ \BBA {} {De Wilde}, L.%
\end{APACrefauthors}%
\unskip\
\newblock
\APACrefYearMonthDay{2025}{}{}.
\newblock
{\BBOQ}\APACrefatitle {{lavaan}: Latent Variable Analysis} {{lavaan}: Latent
  variable analysis}{\BBCQ}\ [\bibcomputersoftwaremanual].
\newblock
\begin{APACrefURL} \url{https://CRAN.R-project.org/package=lavaan}
  \end{APACrefURL}
\newblock
\APACrefnote{R package version 0.6-20}
\newblock
\begin{APACrefDOI} \doi{10.32614/CRAN.package.lavaan} \end{APACrefDOI}
\PrintBackRefs{\CurrentBib}

\bibitem [\protect \citeauthoryear {%
Salehi%
, Rahimzadeh%
, Molaei%
, Zaheri%
\BCBL {}\ \BBA {} Esmaelzadeh-Saeieh%
}{%
Salehi%
\ \protect \BOthers {.}}{%
{\protect \APACyear {2020}}%
}]{%
salehi2020}
\APACinsertmetastar {%
salehi2020}%
\begin{APACrefauthors}%
Salehi, L.%
, Rahimzadeh, M.%
, Molaei, E.%
, Zaheri, H.%
\BCBL {}\ \BBA {} Esmaelzadeh-Saeieh, S.%
\end{APACrefauthors}%
\unskip\
\newblock
\APACrefYearMonthDay{2020}{}{}.
\newblock
{\BBOQ}\APACrefatitle {The relationship among fear and anxiety of {COVID}-19,
  pregnancy experience, and mental health disorder in pregnant women: A
  structural equation model} {The relationship among fear and anxiety of
  {COVID}-19, pregnancy experience, and mental health disorder in pregnant
  women: A structural equation model}.{\BBCQ}
\newblock
\APACjournalVolNumPages{Brain and Behavior}{10}{11}{e01835}.
\newblock
\begin{APACrefDOI} \doi{https://doi.org/10.1002/brb3.1835} \end{APACrefDOI}
\PrintBackRefs{\CurrentBib}

\bibitem [\protect \citeauthoryear {%
Schamberger%
, Schuberth%
\BCBL {}\ \BBA {} Henseler%
}{%
Schamberger%
\ \protect \BOthers {.}}{%
{\protect \APACyear {2023}}%
}]{%
Schamberger2023a}
\APACinsertmetastar {%
Schamberger2023a}%
\begin{APACrefauthors}%
Schamberger, T.%
, Schuberth, F.%
\BCBL {}\ \BBA {} Henseler, J.%
\end{APACrefauthors}%
\unskip\
\newblock
\APACrefYearMonthDay{2023}{}{}.
\newblock
{\BBOQ}\APACrefatitle {Confirmatory composite analysis in human development
  research} {Confirmatory composite analysis in human development
  research}.{\BBCQ}
\newblock
\APACjournalVolNumPages{International Journal of Behavioral
  Development}{47}{1}{89--100}.
\newblock
\begin{APACrefDOI} \doi{10.1177/01650254221117506} \end{APACrefDOI}
\PrintBackRefs{\CurrentBib}

\bibitem [\protect \citeauthoryear {%
Schamberger%
, Schuberth%
\BCBL {}\ \BBA {} Henseler%
}{%
Schamberger%
\ \protect \BOthers {.}}{%
{\protect \APACyear {2025}}%
}]{%
Schamberger2025}
\APACinsertmetastar {%
Schamberger2025}%
\begin{APACrefauthors}%
Schamberger, T.%
, Schuberth, F.%
\BCBL {}\ \BBA {} Henseler, J.%
\end{APACrefauthors}%
\unskip\
\newblock
\APACrefYearMonthDay{2025}{}{}.
\newblock
{\BBOQ}\APACrefatitle {Moderated Mediation with Composites: The composite
  moderated structuralequations approach} {Moderated mediation with composites:
  The composite moderated structuralequations approach}.{\BBCQ}
\newblock
\APACjournalVolNumPages{Behavior Research Methods}{}{}{}.
\newblock
\APACrefnote{under review}
\PrintBackRefs{\CurrentBib}

\bibitem [\protect \citeauthoryear {%
Schuberth%
}{%
Schuberth%
}{%
{\protect \APACyear {2023}}%
}]{%
Schuberth2023HO}
\APACinsertmetastar {%
Schuberth2023HO}%
\begin{APACrefauthors}%
Schuberth, F.%
\end{APACrefauthors}%
\unskip\
\newblock
\APACrefYearMonthDay{2023}{}{}.
\newblock
{\BBOQ}\APACrefatitle {The {H}enseler-{O}gasawara specification of composites
  in structural equation modeling: A tutorial} {The {H}enseler-{O}gasawara
  specification of composites in structural equation modeling: A
  tutorial}.{\BBCQ}
\newblock
\APACjournalVolNumPages{Psychological Methods}{28}{4}{843--859}.
\newblock
\begin{APACrefDOI} \doi{10.1037/met0000432} \end{APACrefDOI}
\PrintBackRefs{\CurrentBib}

\bibitem [\protect \citeauthoryear {%
Schuberth%
, Henseler%
\BCBL {}\ \BBA {} Dijkstra%
}{%
Schuberth%
\ \protect \BOthers {.}}{%
{\protect \APACyear {2018}}%
{\protect \APACexlab {{\protect \BCnt {1}}}}}]{%
Schuberth2018}
\APACinsertmetastar {%
Schuberth2018}%
\begin{APACrefauthors}%
Schuberth, F.%
, Henseler, J.%
\BCBL {}\ \BBA {} Dijkstra, T\BPBI K.%
\end{APACrefauthors}%
\unskip\
\newblock
\APACrefYearMonthDay{2018{\protect \BCnt {1}}}{}{}.
\newblock
{\BBOQ}\APACrefatitle {Confirmatory Composite Analysis} {Confirmatory composite
  analysis}.{\BBCQ}
\newblock
\APACjournalVolNumPages{Frontiers in Psychology}{9}{2541}{}.
\newblock
\begin{APACrefDOI} \doi{10.3389/fpsyg.2018.02541} \end{APACrefDOI}
\PrintBackRefs{\CurrentBib}

\bibitem [\protect \citeauthoryear {%
Schuberth%
, Henseler%
\BCBL {}\ \BBA {} Dijkstra%
}{%
Schuberth%
\ \protect \BOthers {.}}{%
{\protect \APACyear {2018}}%
{\protect \APACexlab {{\protect \BCnt {2}}}}}]{%
Schuberth2018a}
\APACinsertmetastar {%
Schuberth2018a}%
\begin{APACrefauthors}%
Schuberth, F.%
, Henseler, J.%
\BCBL {}\ \BBA {} Dijkstra, T\BPBI K.%
\end{APACrefauthors}%
\unskip\
\newblock
\APACrefYearMonthDay{2018{\protect \BCnt {2}}}{}{}.
\newblock
{\BBOQ}\APACrefatitle {Partial least squares path modeling using ordinal
  categorical indicators} {Partial least squares path modeling using ordinal
  categorical indicators}.{\BBCQ}
\newblock
\APACjournalVolNumPages{Quality \& Quantity}{52}{1}{9--35}.
\newblock
\begin{APACrefDOI} \doi{10.1007/s11135-016-0401-7} \end{APACrefDOI}
\PrintBackRefs{\CurrentBib}

\bibitem [\protect \citeauthoryear {%
Schuberth%
, Schamberger%
, Kemény%
\BCBL {}\ \BBA {} Henseler%
}{%
Schuberth%
\ \protect \BOthers {.}}{%
{\protect \APACyear {2025}}%
}]{%
Schuberth2025}
\APACinsertmetastar {%
Schuberth2025}%
\begin{APACrefauthors}%
Schuberth, F.%
, Schamberger, T.%
, Kemény, I.%
\BCBL {}\ \BBA {} Henseler, J.%
\end{APACrefauthors}%
\unskip\
\newblock
\APACrefYearMonthDay{2025}{}{}.
\newblock
{\BBOQ}\APACrefatitle {The Sum Score Model: Specifying andTesting Equally
  Weighted Composites Using Structural Equation Modeling} {The sum score model:
  Specifying andtesting equally weighted composites using structural equation
  modeling}.{\BBCQ}
\newblock
\APACjournalVolNumPages{Psychometrika}{90}{1}{358--383}.
\newblock
\begin{APACrefDOI} \doi{10.1017/psy.2024.5} \end{APACrefDOI}
\PrintBackRefs{\CurrentBib}

\bibitem [\protect \citeauthoryear {%
Shrout%
\ \BBA {} Bolger%
}{%
Shrout%
\ \BBA {} Bolger%
}{%
{\protect \APACyear {2002}}%
}]{%
Shrout2002}
\APACinsertmetastar {%
Shrout2002}%
\begin{APACrefauthors}%
Shrout, P\BPBI E.%
\BCBT {}\ \BBA {} Bolger, N.%
\end{APACrefauthors}%
\unskip\
\newblock
\APACrefYearMonthDay{2002}{}{}.
\newblock
{\BBOQ}\APACrefatitle {Mediation in experimental and nonexperimental studies:
  New procedures and recommendations.} {Mediation in experimental and
  nonexperimental studies: New procedures and recommendations.}{\BBCQ}
\newblock
\APACjournalVolNumPages{Psychological Methods}{7}{4}{422--445}.
\newblock
\begin{APACrefDOI} \doi{10.1037/1082-989x.7.4.422} \end{APACrefDOI}
\PrintBackRefs{\CurrentBib}

\bibitem [\protect \citeauthoryear {%
Siegel%
, Savalei%
\BCBL {}\ \BBA {} Rhemtulla%
}{%
Siegel%
\ \protect \BOthers {.}}{%
{\protect \APACyear {2025}}%
}]{%
Siegel2025}
\APACinsertmetastar {%
Siegel2025}%
\begin{APACrefauthors}%
Siegel, D.%
, Savalei, V.%
\BCBL {}\ \BBA {} Rhemtulla, M.%
\end{APACrefauthors}%
\unskip\
\newblock
\APACrefYearMonthDay{2025}{}{}.
\newblock
{\BBOQ}\APACrefatitle {Nested Model Comparisons Between Common Factors and
  Composites} {Nested model comparisons between common factors and
  composites}.{\BBCQ}
\newblock
\APACjournalVolNumPages{OSF}{}{}{}.
\newblock
\begin{APACrefURL} \url{https://www.doi.org/10.31234/osf.io/jgwe6\_v2}
  \end{APACrefURL}
\PrintBackRefs{\CurrentBib}

\bibitem [\protect \citeauthoryear {%
Thompson%
}{%
Thompson%
}{%
{\protect \APACyear {1984}}%
}]{%
Thompson1984}
\APACinsertmetastar {%
Thompson1984}%
\begin{APACrefauthors}%
Thompson, B.%
\end{APACrefauthors}%
\unskip\
\newblock
\APACrefYear{1984}.
\newblock
\APACrefbtitle {Canonical correlation analysis: Uses and interpretation}
  {Canonical correlation analysis: Uses and interpretation}.
\newblock
\APACaddressPublisher{Beverly Hills, Calif}{Sage Publications}.
\PrintBackRefs{\CurrentBib}

\bibitem [\protect \citeauthoryear {%
Van~Riel%
, Henseler%
, Kem\'eny%
\BCBL {}\ \BBA {} Sasovova%
}{%
Van~Riel%
\ \protect \BOthers {.}}{%
{\protect \APACyear {2017}}%
}]{%
VanRiel2017}
\APACinsertmetastar {%
VanRiel2017}%
\begin{APACrefauthors}%
Van~Riel, A\BPBI C\BPBI R.%
, Henseler, J.%
, Kem\'eny, I.%
\BCBL {}\ \BBA {} Sasovova, Z.%
\end{APACrefauthors}%
\unskip\
\newblock
\APACrefYearMonthDay{2017}{}{}.
\newblock
{\BBOQ}\APACrefatitle {Estimating hierarchical constructs using consistent
  Partial Least Squares: {T}he case of second order composites of common
  factors} {Estimating hierarchical constructs using consistent partial least
  squares: {T}he case of second order composites of common factors}.{\BBCQ}
\newblock
\APACjournalVolNumPages{Industrial Management \& Data
  Systems}{117}{3}{459--477}.
\newblock
\begin{APACrefDOI} \doi{10.1108/IMDS-07-2016-0286} \end{APACrefDOI}
\PrintBackRefs{\CurrentBib}

\bibitem [\protect \citeauthoryear {%
\v{S}uri\c{n}a%
\ \protect \BOthers {.}}{%
\v{S}uri\c{n}a%
\ \protect \BOthers {.}}{%
{\protect \APACyear {2021}}%
}]{%
Surina2021}
\APACinsertmetastar {%
Surina2021}%
\begin{APACrefauthors}%
\v{S}uri\c{n}a, S.%
, Martinsone, K.%
, Perepjolkina, V.%
, Kolesnikova, J.%
, Vainik, U.%
, Ruža, A.%
\BDBL {}Rancans, E.%
\end{APACrefauthors}%
\unskip\
\newblock
\APACrefYearMonthDay{2021}{}{}.
\newblock
{\BBOQ}\APACrefatitle {Factors related to {COVID-19} preventive behaviors: A
  structural equation model} {Factors related to {COVID-19} preventive
  behaviors: A structural equation model}.{\BBCQ}
\newblock
\APACjournalVolNumPages{Frontiers in Psychology}{12}{}{1--15}.
\newblock
\begin{APACrefDOI} \doi{10.3389/fpsyg.2021.676521} \end{APACrefDOI}
\PrintBackRefs{\CurrentBib}

\bibitem [\protect \citeauthoryear {%
Wang%
, Qiao%
, Li%
\BCBL {}\ \BBA {} Lei%
}{%
Wang%
\ \protect \BOthers {.}}{%
{\protect \APACyear {2021}}%
}]{%
wang_qiao_li_lei_2021}
\APACinsertmetastar {%
wang_qiao_li_lei_2021}%
\begin{APACrefauthors}%
Wang, X.%
, Qiao, Y.%
, Li, W.%
\BCBL {}\ \BBA {} Lei, L.%
\end{APACrefauthors}%
\unskip\
\newblock
\APACrefYearMonthDay{2021}{}{}.
\newblock
{\BBOQ}\APACrefatitle {Parental Phubbing and Children’s Social Withdrawal and
  Aggression: A Moderated Mediation Model of Parenting Behaviors and Parents’
  Gender} {Parental phubbing and children’s social withdrawal and aggression:
  A moderated mediation model of parenting behaviors and parents’
  gender}.{\BBCQ}
\newblock
\APACjournalVolNumPages{Journal of Interpersonal
  Violence}{37}{21-22}{NP19395–NP19419}.
\newblock
\begin{APACrefDOI} \doi{10.1177/08862605211042807} \end{APACrefDOI}
\PrintBackRefs{\CurrentBib}

\bibitem [\protect \citeauthoryear {%
Whitt%
}{%
Whitt%
}{%
{\protect \APACyear {1986}}%
}]{%
Whitt1986}
\APACinsertmetastar {%
Whitt1986}%
\begin{APACrefauthors}%
Whitt, H\BPBI P.%
\end{APACrefauthors}%
\unskip\
\newblock
\APACrefYearMonthDay{1986}{}{}.
\newblock
{\BBOQ}\APACrefatitle {The sheaf coefficient: A simplified and expanded
  approach} {The sheaf coefficient: A simplified and expanded approach}.{\BBCQ}
\newblock
\APACjournalVolNumPages{Social Science Research}{15}{2}{174--189}.
\newblock
\begin{APACrefDOI} \doi{10.1016/0049-089x(86)90014-1} \end{APACrefDOI}
\PrintBackRefs{\CurrentBib}

\bibitem [\protect \citeauthoryear {%
Wold%
}{%
Wold%
}{%
{\protect \APACyear {1982}}%
}]{%
Wold1982c}
\APACinsertmetastar {%
Wold1982c}%
\begin{APACrefauthors}%
Wold, H.%
\end{APACrefauthors}%
\unskip\
\newblock
\APACrefYearMonthDay{1982}{}{}.
\newblock
{\BBOQ}\APACrefatitle {Soft Modeling: {T}he Basic Design and Some Extensions}
  {Soft modeling: {T}he basic design and some extensions}.{\BBCQ}
\newblock
\BIn{} K\BPBI G.~J\"oreskog\ \BBA {} H.~Wold\ (\BEDS), \APACrefbtitle {Systems
  under Indirect Observation: Causality, Structure, Prediction Part II}
  {Systems under indirect observation: Causality, structure, prediction part
  ii}\ (\BPGS\ 1--54).
\newblock
\APACaddressPublisher{Amsterdam}{North-Holland}.
\PrintBackRefs{\CurrentBib}

\bibitem [\protect \citeauthoryear {%
Yu%
, Schuberth%
\BCBL {}\ \BBA {} Henseler%
}{%
Yu%
\ \protect \BOthers {.}}{%
{\protect \APACyear {2023}}%
}]{%
yu_schuberth_henseler_2023}
\APACinsertmetastar {%
yu_schuberth_henseler_2023}%
\begin{APACrefauthors}%
Yu, X.%
, Schuberth, F.%
\BCBL {}\ \BBA {} Henseler, J.%
\end{APACrefauthors}%
\unskip\
\newblock
\APACrefYearMonthDay{2023}{}{}.
\newblock
{\BBOQ}\APACrefatitle {Specifying composites in structural equation modeling: A
  refinement of the \uppercase{H}enseler–\uppercase{O}gasawara Specification}
  {Specifying composites in structural equation modeling: A refinement of the
  \uppercase{H}enseler–\uppercase{O}gasawara specification}.{\BBCQ}
\newblock
\APACjournalVolNumPages{Statistical Analysis and Data Mining}{16}{4}{348--357}.
\newblock
\begin{APACrefDOI} \doi{10.1002/sam.11608} \end{APACrefDOI}
\PrintBackRefs{\CurrentBib}

\bibitem [\protect \citeauthoryear {%
Yu%
, Schuberth%
\BCBL {}\ \BBA {} Henseler%
}{%
Yu%
\ \protect \BOthers {.}}{%
{\protect \APACyear {2025}}%
}]{%
Yu2025}
\APACinsertmetastar {%
Yu2025}%
\begin{APACrefauthors}%
Yu, X.%
, Schuberth, F.%
\BCBL {}\ \BBA {} Henseler, J.%
\end{APACrefauthors}%
\unskip\
\newblock
\APACrefYearMonthDay{2025}{}{}.
\newblock
{\BBOQ}\APACrefatitle {A flexible way to study composites in ecology using
  structural equation modeling} {A flexible way to study composites in ecology
  using structural equation modeling}.{\BBCQ}
\newblock
\APACjournalVolNumPages{Scientific Reports}{15}{1}{}.
\newblock
\begin{APACrefDOI} \doi{10.1038/s41598-025-88675-0} \end{APACrefDOI}
\PrintBackRefs{\CurrentBib}

\bibitem [\protect \citeauthoryear {%
Yu%
, Zaza%
, Schuberth%
\BCBL {}\ \BBA {} Henseler%
}{%
Yu%
\ \protect \BOthers {.}}{%
{\protect \APACyear {2021}}%
}]{%
Yu2021}
\APACinsertmetastar {%
Yu2021}%
\begin{APACrefauthors}%
Yu, X.%
, Zaza, S.%
, Schuberth, F.%
\BCBL {}\ \BBA {} Henseler, J.%
\end{APACrefauthors}%
\unskip\
\newblock
\APACrefYearMonthDay{2021}{}{}.
\newblock
{\BBOQ}\APACrefatitle {Counterpoint: Representing Forged Concepts as Emergent
  Variables Using Composite-Based Structural Equation Modeling} {Counterpoint:
  Representing forged concepts as emergent variables using composite-based
  structural equation modeling}.{\BBCQ}
\newblock
\APACjournalVolNumPages{{SIGMIS} Database}{52}{{SI}}{114--130}.
\newblock
\begin{APACrefDOI} \doi{10.1145/3505639.3505647} \end{APACrefDOI}
\PrintBackRefs{\CurrentBib}

\end{thebibliography}

\end{document}